\documentclass{type_of_file}%

\usepackage{graphicx}
\usepackage{multicol,multirow}
\usepackage{amsmath,amssymb,amsfonts}
\usepackage{mathrsfs}
\usepackage{amsthm}
\usepackage{rotating}
\usepackage{appendix}
\usepackage[backend=biber, style=authoryear, natbib=true, url=true, doi=true]{biblatex}
\usepackage[table]{xcolor} 
\usepackage{colortbl} 
\usepackage{float}
\usepackage{ifpdf}
\usepackage[T1]{fontenc}
\usepackage{newtxtext}
\usepackage{newtxmath}
\usepackage{textcomp}
\usepackage{xcolor}
\usepackage{lipsum}

\theoremstyle{definition}

\numberwithin{equation}{section}
\usepackage{subcaption}
\usepackage[acronym]{glossaries}
\usepackage{pdflscape}
\usepackage{caption}
\makeglossaries
\articletype{ }

\newacronym{aadb}{AADB}{average annual daily bicycle volume}

\newacronym{ml}{ML}{machine learning}

\newacronym{cv}{CV}{cross validation}

\newacronym{logo}{LOGO}{leave-one-group-out}

\newacronym{mae}{MAE}{mean absolute error}

\newacronym{smape}{SMAPE}{symmetric mean absolute percentage error}

\newacronym{gpi}{GPI}{Grouped Permutation Importance}

\newacronym{fs}{FS}{feature selection}

\newacronym{mape}{MAPE}{mean absolute percentage error}

\newacronym{rmse}{RMSE}{root mean squared error}

\begin{document}

\begin{Frontmatter}

\title[Application Paper]{From Counting Stations to City-Wide Estimates:\\
Data-Driven Bicycle Volume Extrapolation}

\author[1]{Silke K.~Kaiser*}
\author[2]{Nadja Klein}
\author[1]{Lynn H.~Kaack}

\authormark{Kaiser \textit{et al}.}

\address[1]{\orgdiv{Data Science Lab}, \orgname{Hertie School}, \orgaddress{\city{Berlin}, \postcode{10117},   \country{Germany}}}

\address[2]{\orgdiv{Scientific Computing Center}, \orgname{Karlsruhe Institute of Technology}, \orgaddress{\city{Karlsruhe}, \postcode{76131},  \country{Germany}}.\par *Corresponding author. \email{s.kaiser@phd.hertie-school.org}}

\authormark{Kaiser et al.}


\abstract{S\begin{abstract}
Shifting to cycling in urban areas reduces greenhouse gas emissions and improves public health. Street-level bicycle volume information would aid cities in planning targeted infrastructure improvements to encourage cycling and provide civil society with evidence to advocate for cyclists' needs. Yet, the data currently available to cities and citizens often only comes from sparsely located counting stations. This paper extrapolates bicycle volume beyond these few locations to estimate bicycle volume for the entire city of Berlin. We predict daily and average annual daily street-level bicycle volumes using machine-learning techniques and various public data sources. These include app-based crowdsourced data, infrastructure, bike-sharing, motorized traffic, socioeconomic indicators, weather, and holiday data. Our analysis reveals that the best-performing model is XGBoost, and crowdsourced cycling and infrastructure data are most important for the prediction. We further simulate how collecting short-term counts at predicted locations improves performance. By providing ten days of such sample counts for each predicted location to the model, we are able to halve the error and greatly reduce the variability in performance among predicted locations.

\end{abstract}

}

\end{Frontmatter}

\section{Introduction}


Shifting from motorized transport to bicycles improves cardiorespiratory health, reduces the risk of cancer mortality \parencite{oja_health_2011, woodcock_public_2009} and reduces greenhouse gas emissions \parencite{h-o_portner_dc_roberts_es_poloczanska_k_mintenbeck_m_tignor_a_alegria_m_craig_s_langsdorf_s_loschke_v_moller_a_okem_eds_ipcc_2022}.
A promising lever to encourage people to cycle in cities is infrastructure improvements:  Previous studies have shown that adult cyclists, and especially, women, prefer to ride infrastructure specifically designated for them \parencite{dill_bicycling_2009, garrard_promoting_2008}. This is probably because a cyclist is less likely to be involved in an accident when riding in a separate bicycle lane \parencite{morrison_-road_2019}; noting that most cyclists perceive risks in accordance with actual risk \parencite{moller_cyclists_2008}.  However, introducing new bike lanes is expensive and often highly contested due to limited resources, such as funding and road space. Thus, data-driven approaches are crucial for accurately targeting infrastructure improvements to areas with the greatest need \parencite{olmos_data_2020,larsen_build_2013}. \\


One relevant piece of information for such data-driven approaches is bicycle volume data. Currently, most of this data is collected by permanently installed bicycle counting stations, providing information on cyclists passing by a specific location. Due to their high cost, these stations are sparsely located across a road network. At the same time, several data sources related to cycling are openly available \parencite{romanillos_big_2016}. Given the scarcity of bicycle volume data on the one hand and the abundance of related data on the other hand, clamors for methods that are able to make use of all available information in order to better predict bicycle volumes at a fine-grained scale. We address this by combining \acrlong{ml} (\acrshort{ml}) methods with a wide variety of available data sources to extrapolate bicycle volume to a much higher spatial resolution.
With this machinery, we aim to answer three important questions. First, can we predict bicycle volume at unseen locations using a variety of data? Second, which of these data sources are the most relevant for prediction? And third, how much can the performance be improved by adding sample counts for the predicted locations? \\

Researchers have identified several datasets related to bicycle volume that have proven useful, especially for interpolating missing observations in bicycle count data. These include data sources that have long been available, such as weather, holiday, infrastructure, and socioeconomic indicators \parencite{miranda-moreno_weather_2011, strauss_spatial_2013, holmgren_prediction_2017}. The potential of additional available data sources associated with the widespread use of smartphones has also been explored \parencite{lee_emerging_2020}. This includes valuable information from crowdsourced bicycle usage data, in particular, from the Strava application \parencite{lee_strava_2021, kwigizile_leveraging_2022}, bike-sharing protocols \parencite{miah_estimation_2022} or the use of photos and tweets \parencite{wu_photos_2017}. \\

Among available studies, some extrapolate bicycle volume using only a few of these data sources. For instance, \cite{miah_challenges_2022} explore how counting station data can be merged with crowdsourced data to estimate bicycle volumes across street networks using clustering and nonparametric modeling. They find that relying solely on crowdsourced data as an additional input to counts is challenging, particularly due to oversampling from counting stations located at high-volume locations. Similar studies estimate cyclists' exposure employing various data sources and using classical regression approaches \parencite{sanders_ballpark_2017, griswold_pilot_2011}, mixed effects models \parencite{dadashova_random_2020} or  Poisson regressions \parencite{roy_correcting_2019}. 
In addition to traditional statistical approaches, ML methods have been increasingly applied over the past decade. For instance, \parencite{sekula_estimating_2018, das_interpretable_2020,zahedian_estimating_2020} have proven how ML methods can be leveraged for the extrapolation of motorized traffic. However, to the best of our knowledge, there is no study that combines ML methods with a large variety of different data sources to provide reliable, fine-grained predictions of bicycle counts beyond available counting stations. \\

Our paper showcases our approach in the city of Berlin. In Germany's largest city, with 3.6 million residents, the transportation mode share for walking and cycling lies at 37\%, aligning with the European average of 42\% \parencite{european_metropolitan_transport_authorities_barometer_2021}, making it a suitable representative case for our analysis. 
We implement and compare different ML algorithms to predict the daily and \acrlong{aadb} (\acrshort{aadb}) at unseen locations. We use a wide array of features, many of which have proven pertinent in previous studies (see Table \ref{table:overview_factors_to_predict_cycling_volume} for an overview). To identify the most relevant data sources, we perform a grouped base permutation feature importance. 
In addition, to further improve the predictions, we evaluate whether collecting sample counts at unseen locations would be purposeful and what is the best strategy to collect this data.

\begin{table}[H]
\tabcolsep=0pt%
\TBL{\caption{Overview of data types used in this paper to predict bicycle volume and their use in other publications: including  crowdsourced (Strava)[Crowds.], infrastructure [Infr.], weather [Weath.], socioeconomic [Socio.], bike-sharing [B.-S.], public and school holidays [Hol.] and motorized traffic [Moto.].\label{table:overview_factors_to_predict_cycling_volume}}}

    \begin{tabular*}{\textwidth}{@{\extracolsep{\fill}}lccccccc@{}}\toprule%
    Reference / Data Source & Crowds. & Infr. & Weath. & Socio. & B.-S & Hol.&Moto. \\\midrule
    \cite{miah_estimation_2022}         & \checkmark       & \checkmark & \checkmark    & \checkmark &  \checkmark   &   &\\
    \cite{dadashova_random_2020}        & \checkmark       & \checkmark & \checkmark    & \checkmark &      &   &\\
    \cite{kwigizile_leveraging_2022}    & \checkmark       & \checkmark &\checkmark     & \checkmark &      &   &\\ 
    \cite{hochmair_estimating_2019}     & \checkmark       & \checkmark &      & \checkmark &      &   &\\
    \cite{sanders_ballpark_2017}        & \checkmark       & \checkmark &      & \checkmark &      &   &\\
    \cite{nelson_generalized_2021}      & \checkmark       &   & \checkmark    & \checkmark &      &   &\\
    \cite{hankey_facility-demand_2016}  &         & \checkmark & \checkmark    & \checkmark &      &   &\\ 
    \cite{strauss_spatial_2013}         &         & \checkmark & \checkmark    & \checkmark &      &   &\\  
    \cite{roy_correcting_2019}          & \checkmark       & \checkmark &      &   &      &   &\\
    \cite{el_esawey_daily_2018}         &         &   & \checkmark    &   &      &   &\\ 
    \cite{holmgren_prediction_2017}     &         &   & \checkmark    &   &      & \checkmark & \\
    \cite{miranda-moreno_weather_2011}  &         &   &\checkmark     &   &      &   &\\
    \cite{lu_designing_2017}            &         & \checkmark &      &   &      &   &\\
    \hline
    This paper &\checkmark &\checkmark &\checkmark &\checkmark & \checkmark&\checkmark &\checkmark  \\
    \botrule
    \end{tabular*}%

\end{table}


\section{Results}

\subsection{Data sources}

Our study uses data from 20 long-term bicycle counting stations in Berlin, which continuously measure the number of passing bicycles per hour. In addition, we employ data from 12 short-term counting stations, where counts are conducted on individual days throughout the year \parencite{senate_department_for_the_environment_mobility_consumer_and_climate_protection_berlin_jahresdatei_2022}. 
To accurately predict bicycle counts, we make use of information contained in a variety of further sources. These include data on infrastructure, socioeconomic factors, motorized traffic, weather, holidays, bike-sharing, and from a crowdsourcing application that tracks cyclists (Strava application). Bike-sharing and Strava data directly represent bicycle traffic. But they attract different users and differ in the type of information they provide. The former describes the exact time and origin-destination-pairs of individual trips taken on short-term free-floating rented bikes. The latter are anonymized georeferenced data from an application, which are aggregated to provide the number of trips for a region and for road segments between intersections based on tracking users as they ride. The bike-sharing, crowdsourced, and motorized traffic data are feature-engineered, to indicate the usage volume, respectively of passing motorized traffic within different radii around counting stations. The socioeconomic and infrastructure features are assigned in accordance with the location of the counting stations. Further details on the distinct data sources, including data clearing and feature engineering are provided in the  Methods Section \ref{sec:methods_data_description}. A list of all features is provided in Table \ref{table:brief_overview_features} together with references to the data sources. The bike-sharing data is only available for April to December 2019 and June to December 2022. Therefore, we set our study period to these periods. This also largely omits the period of the COVID-19 pandemic and its impact on transportation.

\begin{table}
    \centering
    \renewcommand{\arraystretch}{1.5} 
        \caption{Overview of the features per data source used in this study.}
    \label{table:brief_overview_features}
    
    \begin{tabular}{p{2.1cm}|p{7.2cm}|p{1cm}|p{2.7cm}}
    \toprule
         Data category      &Description of features & No of\hspace{0.2cm}  features& Data source\\
         \midrule
          Crowdsourced             & Number of trips originating, arriving, or happening; with respect to leisure and commute, with respect to different times of the day, with respect to the weekend, with respect to different personal characteristics (age, sex), with respect to normal and e-bikes, as well as average speed. Both for hexagon and street segment data &91& \cite{strava_metro_strava_2023}\\
           Infrastructure     & Latitude, longitude, distance to city center, maximum speed, bicycle lane type, number of shops/education centers/hotels/hospitals for various radii, percent of area used for farming/horticulture/cemeteries/waterways/industry/ private gardening/parks/traffic areas/forests/ residential housing &31& \cite{openstreetmap_contributors_planet_2017, senate_department_for_urban_development_building_and_housing_lebensweltlich_2023}\\
            Weather            & Average/maximum/minimum temperature, precipitation, maximum snow depth, sunshine duration, wind speed, wind direction, peak wind gust, dew point, air pressure, humidity  & 10& \cite{meteostat_weathers_2022}\\
         Socioeconomic      & Population density, total number of inhabitants, average age, gender distribution,  share of population with migration background, share of foreigners, share of unemployed, share of population with tenure exceeding 5 years, rate moving to/from area, age-specific demographic proportions, greying index, birth rate &15& \cite{senate_department_for_urban_development_building_and_housing_lebensweltlich_2023, berlin-brandenburg_office_of_statistics_kommunalatlas_2020}\\
         Bike-sharing       & Number of bicycles originated, returned rented within various radii&18& \cite{citylab_berlin_shared_2019, nextbike_official_2020, kaiser_bike-sharing_2023}\\
         Holiday            & School holiday, public holiday &2& \cite{senate_department_for_education_youth_and_family_ferientermine_2022}\\
         Motorized traffic & total number and speed of vehicles/cars/lorries within different radii  &12& \cite{berlin_open_data_verkehrsdetektion_2022} \\
        Time               & Month, day of month, weekday, weekend, year, &5 &  inherent\\
         \midrule
         Total number of features & &184 & \\
         \bottomrule
    \end{tabular}

\end{table}

\subsection{Spatial extrapolation using multi-source data}
We train our model using data from existing counting stations as ground truth, iteratively omitting one counter from the training dataset, and evaluating the performance for this omitted location. We compare the performance of different ML algorithms on this task. A description of the models, the feature selection, and the hyperparameter tuning can be found in the Methods Section \ref{sec:data_analysis}. 
We evaluate the predictions at the daily and at the annual scales. The daily scale is valuable for providing a more detailed picture of the variation throughout the year, and it is relevant for understanding the effects of intra-week variation, special events, and seasonal weather conditions \parencite{yi_inferencing_2021, sekula_estimating_2018, zahedian_estimating_2020}. For infrastructure planning decisions, annual averages may be sufficient. The \acrshort{aadb} is the average number of bicycles that pass a given location per day for a given year. We compute the performance for the \acrshort{aadb} by predicting the daily counts and evaluating their average against the annual ground truth average. Since the counting station data is recorded hourly, we sum up the measurements for each day to obtain daily measurements.
To simulate extrapolation, we evaluate our models using \acrlong{logo} (\acrshort{logo}) \acrlong{cv} (\acrshort{cv}). The method follows the same principle as standard \acrshort{cv} but differs in how the data is partitioned. Instead of random partitioning, the data is organized into distinct groups, which, in our case, correspond to counting stations. Consequently, the model is trained on observations from all but one long-term counting station and then evaluated on this hold-out long-term counting station.

In addition, we use the each short-term counting location as test data for a model trained on all long-term counting stations. We provide the average error across stations, which implies that each location is equally weighted in the test data. 
When computing these predictions, it is important to note that the hourly long-term data are measured from 0h-24h, while the short-term counts only from 7h-19h. Hence, we train the model, predicting the short-term locations, only on daily measurements, which are computed as the sum of the 7h-19h hourly measurements. We also perform the analysis of the long-term stations on daily measurements based on 0h-24h and 07-19h data separately. The former allows us to infer day effects for long-term stations, and the latter can be used to compare results with the short-term counting predictions. 
In order to provide information on the absolute and relative size of our errors, we use the \acrlong{mae} (\acrshort{mae}), and the \acrlong{smape} (\acrshort{smape}) as evaluation metrics and train the models on various ML algorithms (see Methods Section \ref{sec:data_analysis}). Additionally, we include a baseline, in which we use the mean across the observations in training data as the prediction.\\

\begin{table}
      \caption{Errors for the various machine learning models at the daily, and average annual daily bicycle volume (\acrshort{aadb}) scale. The gray background implicates the columns employed as the criterion for model selection. }%
  \label{Table:Results_one_city_model}%
\small
  \centering
  \begin{subtable}{\linewidth}
    \centering \caption{MAE}

    \begin{tabular}{l>{\columncolor[gray]{0.8}}rrr|>{\columncolor[gray]{0.8}}rrr}
        \toprule
        Dimension       & daily & daily & daily & AADB & AADB & AADB \\
        Time              &  0h-24h & 7h-19h &   7h-19h &  all day & 7h-19h &   7h-19h \\
        Counter type      &long-term &long-term &short-term&long-term &long-term &short-term\\
        Evaluation        &  LOGO &  LOGO &  test &  LOGO &  LOGO &  test \\
        \cline{2-7}
                          &  (1) & (2) &  (3)&  (4) & (5) &  (6)\\
        \midrule
            Linear regression       &5389.92& 3559.30 &2628.23                  &6617.89 &3841.38& 2478.11\\
            Decision tree           &1966.29& 1466.07 &1189.90                  &1715.7 &\textbf{1263.26} &1242.39\\
            Random forest           &1718.33 &1421.11 &\textbf{1008.36}&\textbf{1581.26}& 1310.36 &1067.89\\
            Gradient boosting       &1769.75& 1526.39 &1601.57                  &1692.93 &1463.56& 1488.24\\
            XGBoost                 & \textbf{1634.61} &\textbf{1336.16} &1165.29        & 1557.19& 1293.49 &1182.39\\
            Support vector machine                     &1773.70 &1500.65 &1078.28         & 1653.02& 1386.09& \textbf{899.02}\\
            Shallow neural network & 2107.05 &1691.18 &1674.34& 1985.46 &1588.54 &1536.93\\
            \midrule
            Baseline& 2400.15 &1947.79 &1683.71 &2400.15 &1947.79 &1683.71\\
        \bottomrule
        \end{tabular}
    \label{subtab:a}
  \end{subtable}%
  
  \begin{subtable}{\linewidth}
    \centering  \caption{SMAPE}
    \begin{tabular}{lrrr|rrr}
        \toprule
        Dimension         & daily & daily & daily & AADB & AADB & AADB \\
        Time              &  0h-24h & 7h-19h &   7h-19h &  0h-24h & 7h-19h &   7h-19h \\
        Counter type      &long-term &long-term &short-term&long-term &long-term &short-term\\
        Evaluation        &  LOGO &  LOGO &  test &  LOGO &  LOGO &  test \\
        \cline{2-7}
                          &  (1) & (2) &  (3)&  (4) & (5) &  (6)\\
        \midrule
        Linear regression &127.92 &117.68& 95.80            &126.83 &114.11& 86.79\\
        Decision tree &51.72 &51.85 &47.50                  &44.51 &47.10& \textbf{43.73}\\
        Random forest &46.04 &46.47 &\textbf{46.97}         &41.63 &\textbf{42.88} &48.02\\
        Gradient boosting &43.21 &47.14 &64.81              &40.01 &44.08 &59.04\\
        XGBoost &\textbf{41.24} &\textbf{46.67} &70.64      &\textbf{38.86}& 44.27& 67.38\\
        Support vector machine  &47.06 &48.12 &55.38                &44.85& 43.88 &47.70\\
        Shallow neural network & 57.22& 56.99 &70.10 &52.86& 53.22 &63.09 \\
        \midrule
        Baseline &66.62 &65.67 &70.62 &66.62& 65.67& 70.62\\
        \bottomrule
        \end{tabular}
    \label{subtab:b}
  \end{subtable}
\end{table}

\begin{figure}[b]
  \begin{subfigure}{7cm}
    \centering
    \includegraphics[width=\textwidth]{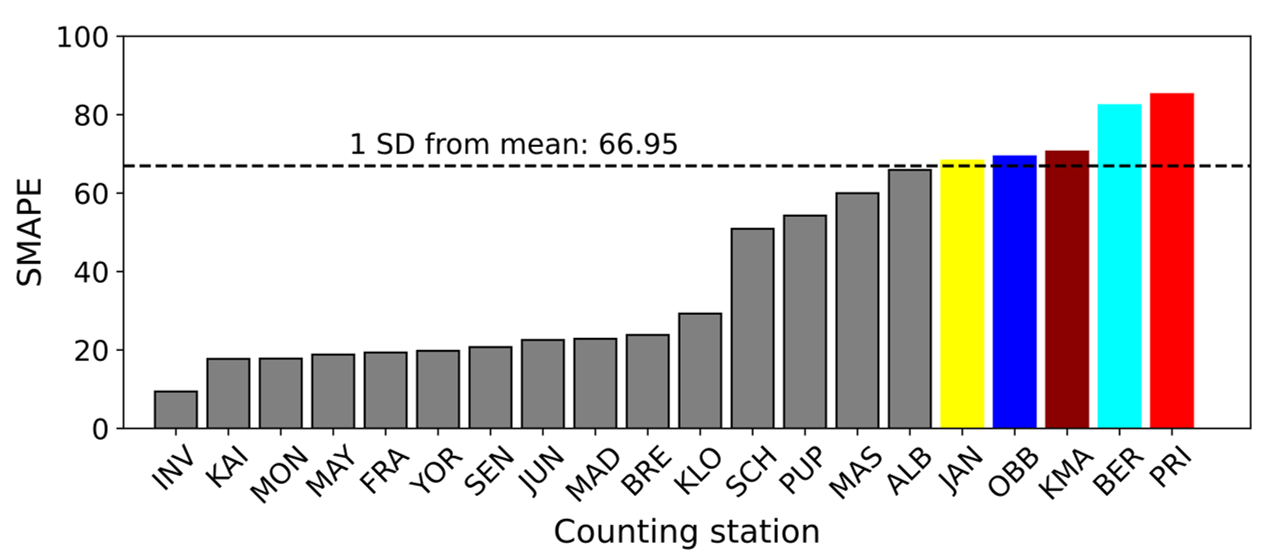}
        \caption{\textit{Model using wide array of inputs (daily)}}
    \label{figure:in_depth_SMAPE_a}
  \end{subfigure}
    \begin{subfigure}{7cm}
    \centering
    \includegraphics[width=\textwidth]{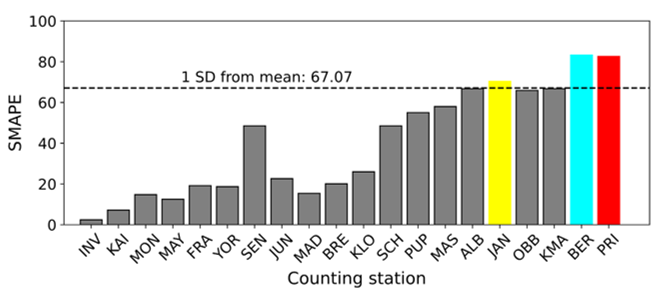}
        \caption{\textit{Model using wide array of inputs (\acrshort{aadb})}}
    \label{figure:in_depth_SMAPE_b}
  \end{subfigure}
      \begin{subfigure}{7cm}
    \centering
    \includegraphics[width=\textwidth]{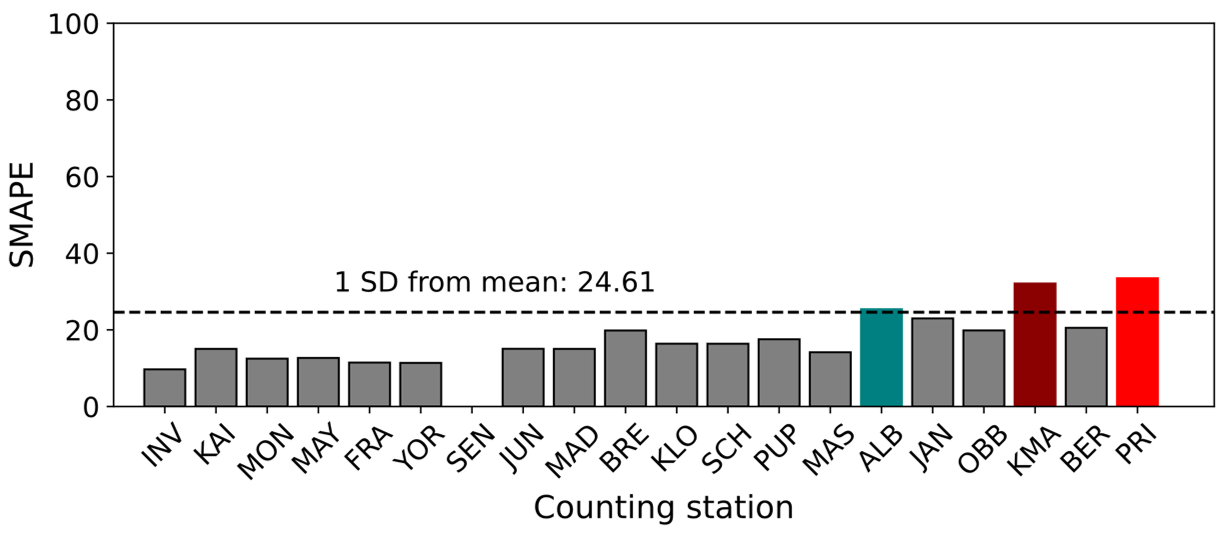}
        \caption{\textit{Model using a wide array of inputs and ten days of sample data (daily)}}
    \label{figure:in_depth_SMAPE_c}
  \end{subfigure}
          \begin{subfigure}{7cm}
    \centering
    \includegraphics[width=\textwidth]{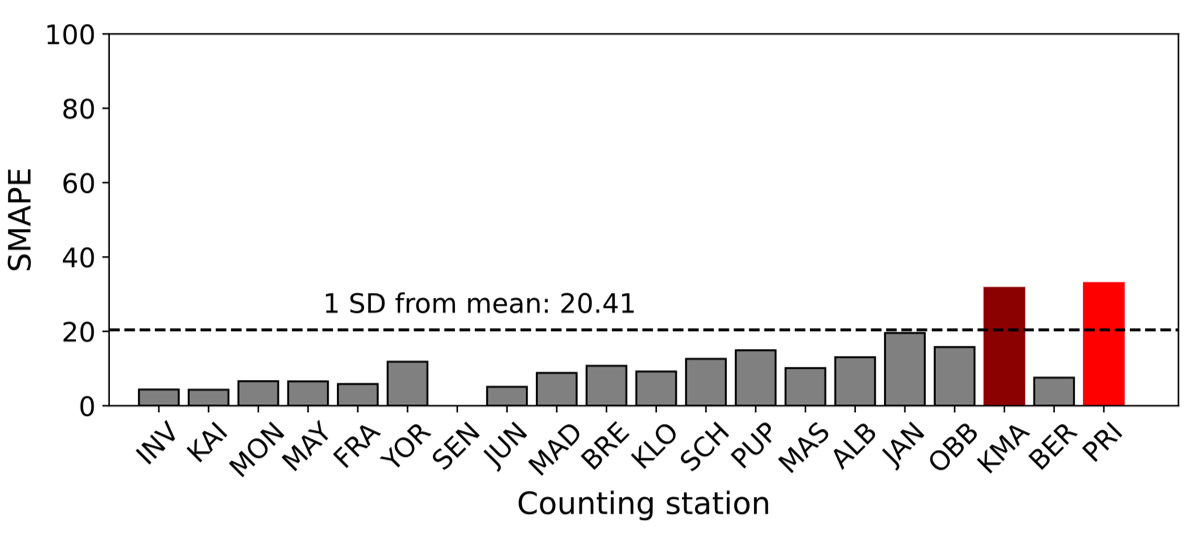}
        \caption{\textit{Model using a wide array of inputs and ten days of sample data (\acrshort{aadb})}}
    \label{figure:in_depth_SMAPE_d}
  \end{subfigure}
\caption{\textit{Performance of XGBoost model at the daily level and for average annual daily bicycle volume estimations (\acrshort{aadb}) across the individual counting stations. Subfigure b) and d) were trained on ten days' worth of sample data and on the additional long-term counting stations (full-city model specification). Highlighted in all graphs are the counting stations whose error exceeds or is below a deviation of 1 standard deviation from the mean. The color coding and the ordering of the counting stations across all subplots are the same to ensure comparability. The counting station 'SEN' is left out in subplot b) and d), due to the small number of observations available.}}
\label{fig:histograms_over_counters}
\end{figure}

We find that ensemble methods (XGBoost, gradient boosting, random forest, and decision trees) outperform the baseline, support vector machines, linear regression, and shallow neural networks (Table \ref{Table:Results_one_city_model}). 
We select XGBoost as the best performing model based on the \acrshort{logo} analysis with all long-term counting stations on the 0h-24h data as evaluated by the MAE. We note that this model does not produce the lowest errors when predicting the short-term counts, with a larger discrepancy in the \acrshort{smape} compared to the \acrshort{mae}. 
To analyze the performance of the XGBoost model in more detail, we looked into the variation of \acrshort{smape} between stations. At the daily scale, the model performs quite well for more than half the stations (\acrshort{smape} of around 20), while for some the \acrshort{smape} exceeds 80 (Figures \ref{figure:in_depth_SMAPE_a}), and the performance also varies considerably between counters for the AADB (Figure \ref{figure:in_depth_SMAPE_c}). The poorly performing locations each have a high variance in their measurements, and each of these locations is either consistently over-predicted or under-predicted. Our analysis revealed no further common characteristics of the worst-performing counters that would allow us to pinpoint where the model is failing. We conclude that there are latent factors within the data generation process that remain unaccounted for, despite our comprehensive inclusion of a wide range of features from the existing literature. We will explore how this can be mitigated using sample counts in Section \ref{sec:sample_counts}.

\subsection{Relative importance of input data sources}

Each data source used requires time and effort for acquisition, cleaning, and integration. Given the variety of sources used in this study, we explore their relative importance so that city officials considering a similar modeling approach can anticipate which ones are essential to obtain.   \\

Feature importance measures the contribution of the feature to the prediction of the target variable.
Given the large number of employed features, which can be grouped by their data source (Table \ref{table:brief_overview_features}), we choose to evaluate their grouped importance. We compute the grouped feature importance at the daily scale, using the \acrlong{gpi} \parencite[\acrshort{gpi};][]{plagwitz_supporting_2022}, which is described in detail in Section \ref{sec:data_analysis}. 
Additionally, we focus on the \acrshort{smape} error, as correctly predicting both relatively busy and relatively slow roads is valuable when deciding where to prioritize infrastructure. Finally, since we want to get a comprehensive picture of the daily traffic situation, including at night, we use the data for the 0h-24h time window. We train the model on all long-term counting stations. Within \acrshort{gpi}, we compute 100 permutations and use repeated  $5$-fold stratified \acrshort{cv}. \\

The \acrshort{gpi} reveals that crowdsourced Strava application data is the most important group, followed by time, infrastructure, and socioeconomic indicators (Figure \ref{figure:feature_importance_SMAPE}). The crowdsourced information is much more relevant than the bike-sharing data. Wile both directly represent bicycle traffic, the movement patterns of individuals tracking their trips turn out to be more indicative of the overall cycling volume of people renting short-term bikes. Therefore, consistent with previous research, we find that Strava indicators are very useful for estimating cycling volumes \parencite{sanders_ballpark_2017,hochmair_estimating_2019, kwigizile_leveraging_2022}.

\subsection{Proof of concept of multi-source model }

\begin{figure}[h]%
      \begin{subfigure}{7cm}
        \centering
            \includegraphics[width=0.9\textwidth]{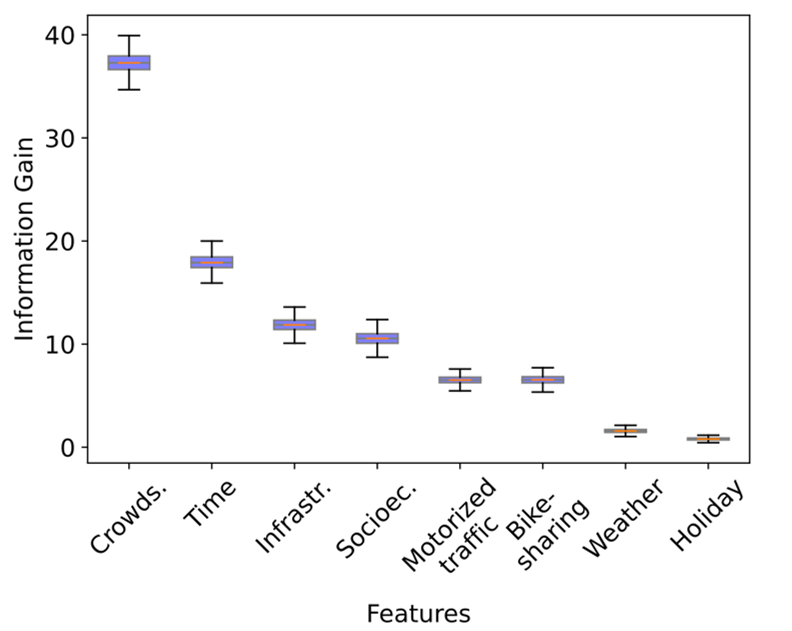}
            \caption{\textit{Information gain computed via a grouped permutation importance for the XGBoost model at the daily level using \acrshort{smape}. All features were assigned to a group (see Table \ref{table:brief_overview_features}).}}
        \label{figure:feature_importance_SMAPE}
      \end{subfigure}
       \hfill 
        \begin{subfigure}{7cm}
        \centering
        \includegraphics[width=0.9\textwidth]{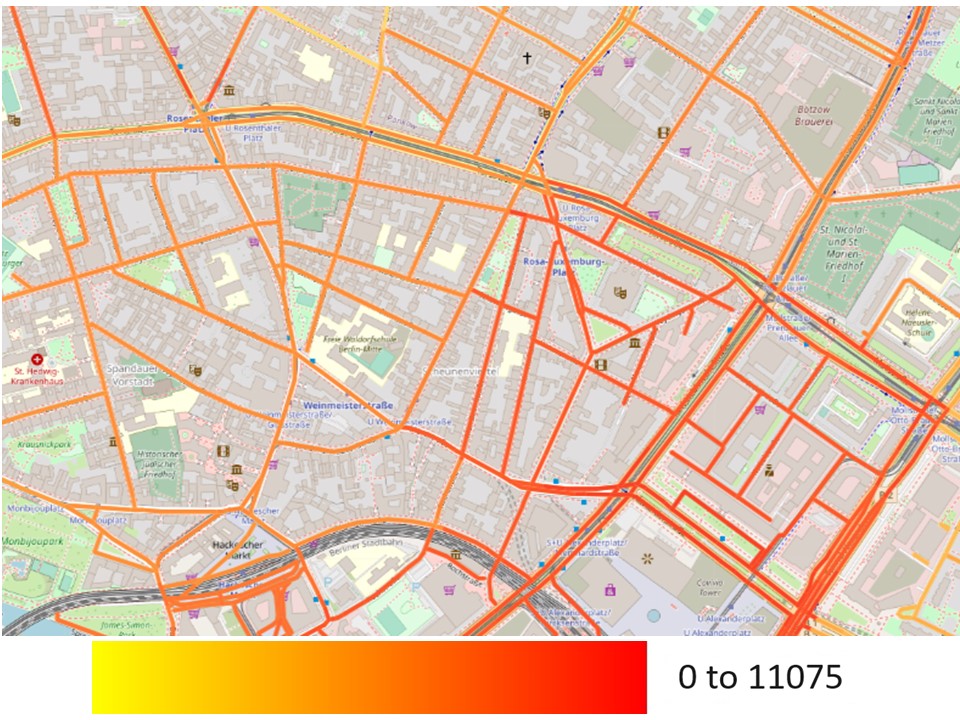}
            \caption{\textit{Proof of concept: Application of the XGBoost model to a subset of Berlin streets to predict the streetwise daily bicycle volume. The prediction in the picture is for 20.09.2022.}}
            \label{fig:proof_pf_concept_streetwise}
        \label{figure:proof_of_concept}
      \end{subfigure}
    \caption{\textit{Feature importance and proof of concept based on an XGBoost model trained on data of all available long-term counting stations.}}
    \label{fig:proof_of_conept_and_feature_importances}
\end{figure}

We empirically demonstrate benefits from our multi-source model (XGBoost trained on all available long-term counting stations) by simulating daily streetwise bicycle volume in a subarea of Berlin for the month of September 2022. Figure \ref{fig:proof_pf_concept_streetwise} shows a snapshot from the simulation, which is available online at \href{https://silkekaiser.github.io/research}{https://silkekaiser.github.io/research}.
We predict streetwise bicycle volume for Berlin. Precisely, we predict volume for every street segment between two intersections. Since our estimates are based on discrete point locations, we compute the midpoint for each street segment and base our estimates on these points. \\

We find that the demonstration effectively captures temporal variations, especially between weekends and weekdays. However, the spatial aspects of the predictions could be more convincing. The model often predicts that adjacent streets have similar bicycle volumes and fails to detect high outliers. This shortcoming is likely due to the construction of features based on large radii. Nevertheless, the model reasonably captures the differences between major streets and residential areas, picking out high and low-traffic zones.

\subsection{Spatial extrapolation using additional sample count data \label{sec:sample_counts}}

Our multi-source model has only a limited ability to reproduce spatial patterns of cycling volume. Here, we investigate whether collecting additional location-specific bicycle volume sample counts improves the predictive performance at unseen locations on a daily scale and what is the most effective strategy for conducting them.\\

\cite{hankey_day--year_2014} elaborate on the usefulness of short-term counts to estimate annual averages for non-motorized traffic using scaling factors. They find that as the number of observation days increases, the extrapolation error decreases, but that the incremental gains become modest after the first week. Also, the advantage derived from conducting counts on consecutive days is minimal compared to nonconsecutive days. We seek to revisit their findings in the context of  \acrshort{ml}. We chose to simulate three different sample data collection strategies: Firstly, the collection of data is commissioned for one day at a time (1-day). The days are selected at random throughout the year. In the second and third strategies, we simulate the collection of data on three (3-day) or seven (7-day) consecutive days. Also, these multi-day periods are randomly distributed throughout the year. We compare the performance of the model with data from each of those three different sampling strategies. We simulate a collection of up to 28 days.\\

We simulate this using 19 long-term counting stations only, as all short and one long-term station have too few observations per location available. We use the XGBoost model with \acrshort{smape}. As before, we evaluate the performance by iterating over the counting stations. Each counting station serves once as the new (``hold-out'') location. For that location, we randomly choose some of the available data to represent sample counts performed at that location, following the three sampling strategies (1, 3 or 7-day). We use the remaining data from the location as the test set. 
For training, we implement two scenarios. For the first scenario, we make use of all available data and train the model on both the sample data and the data from the other counting stations. We give a weight of 25\% to the sample counts and 75\% to the other counting stations' data. Please refer to the Methods Section \ref{sec:data_analysis} for details on the weights. This ``full-city'' scenario benefits from both location-specific sample data and city-wide long-term information.  
For the second scenario, we train the model on the sample data only. Since it only uses information from the location in question, we refer to this model as the ``location-specific'' scenario. Thus, by definition, the training data for this model exhibits no variation in infrastructure and socioeconomic features, as these features only vary across locations. We then use both scenario models to perform prediction on the test set. We repeat this process for each counting station and compute the average across the resulting errors. This procedure is repeated 10 times with different sample days, to allow for uncertainty estimation and provide 95\% confidence intervals. We train and evaluate the models after each additional day of data collection. This allows a comparison of the different approaches for as little as 1 day and as much as 28 days of additionally collected data. 
As a simple baseline, we include the error of predicting the site-specific volume as the mean of the sample data collected at the respective location. 
\\

\begin{figure}[t]
  \begin{subfigure}{7cm}
    \includegraphics[width=\textwidth]{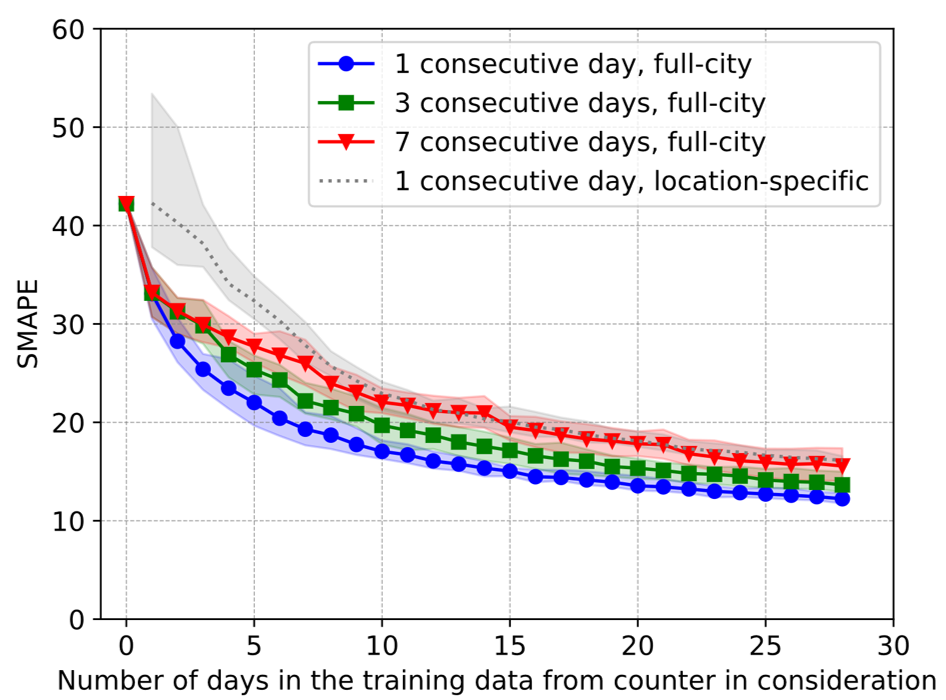}
    \caption{\textit{Combined sample and long-term data for training (full-city)}}
    \label{figure:SMAPE_across_locations_without_sample}
  \end{subfigure}
  \begin{subfigure}{7cm}
    \centering\includegraphics[width=\textwidth]{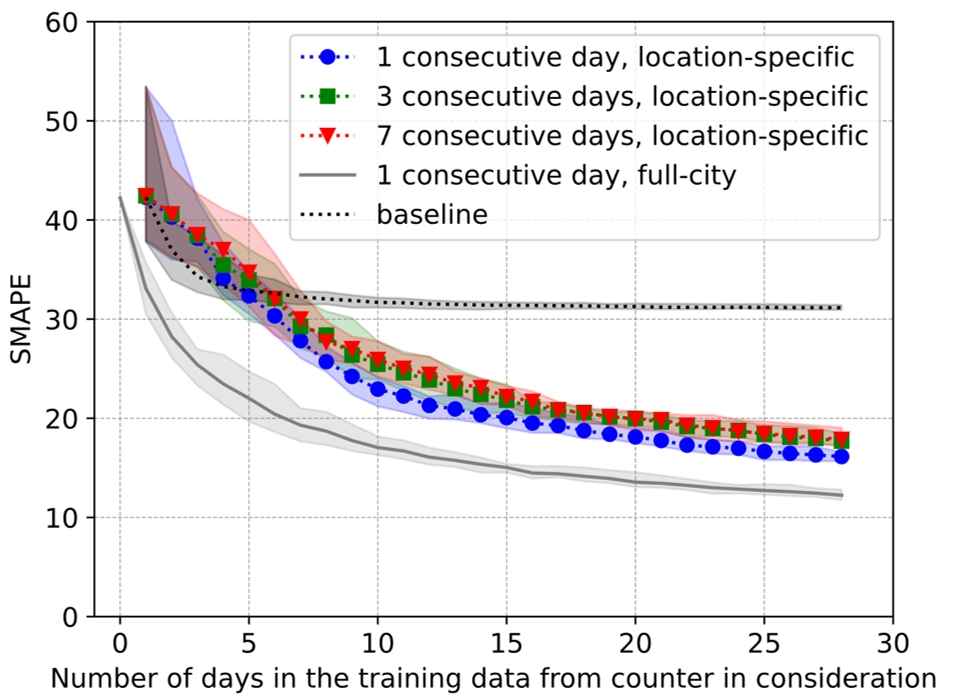}
    \caption{\textit{Using only sample data for training (location-specific)}}
    \label{figure:SMAPE_across_locations_with_sample}
  \end{subfigure}
\caption{\textit{Shown is the effect of collecting additional sample data at a new location to predict the daily volume of bicycles using XGBoost. In the left diagram, the models are trained on the full-city available data, both long-term data from other sites and sample data from the location in question; in the right diagram, the models are trained on location-specific sample data only. Best-performing specifications are depicted in gray in the other plot to allow for comparison. The error is the average over the 19 counting stations used, with 95\% confidence intervals calculated from 10 repeated samples. }}
\label{fig:sample_data_collection}
\end{figure}

 
Sample data collection notably enhances predictive performance for new locations in the full-city scenario (Figure \ref{fig:sample_data_collection}). In the location-specific scenario, two or more days' worth of sample data already outperforms a model without any location-specific data.
Sampling only one day at a time is the superior collection strategy, and this advantage is more pronounced for the full-city scenario. Collecting data on as many different days as possible may provide an advantage, as seasonal effects are better captured. This can also be seen in the curves for the 7-day and 3-day strategies, where the error decreases after the 7th and 14th as well as after the 3rd and 6th day when new random dates for the collection periods are chosen (\ref{figure:SMAPE_across_locations_without_sample}). Given that setting up counting infrastructure at new locations may be costly, the 3-day and 7-day approaches may still yield sufficient results at lower costs. 
Moreover, we find that the full-city approach using the 1-day strategy consistently outperforms the location-specific approaches. A comparison of the 1-day strategy between the two approaches shows that to achieve a \acrshort{smape} of 20, one would need to collect on average 7 days of sample data using the full-city scenario or 14 days using the location-specific model. This underscores the fact that models can benefit greatly from information obtained at locations other than the one under consideration. Finally, we find that the use of multi-source data is also highly relevant when using sample data, and simple averages over the counts do not suffice. The baseline error never drops below 30, while the errors for the location-specific models are below 20 after 25 days of sample counts (Figure \ref{figure:SMAPE_across_locations_with_sample}). This demonstrates the importance of leveraging multi-source data in combination with sample counts.\\

Based on these results, we seek to provide a numerical comparison of the performance of a model with sample data to the simple multi-source model. We train an XGBoost model on the full-city scenario, in combination with multi-source data and ten days' worth of sample counts using the 1-day strategy. We collect these ten days randomly across all observations and across both years. Again, to account for the randomness in the sample data collection, we compute 10 repeated samples and take their average. With this approach, we can predict new locations at the daily scale with an average \acrshort{smape} of 17.44 (in comparison to the multi-source model of 41.24) and an \acrshort{mae} of 594.59 (1634.61). For the \acrshort{aadb}, we get 11.94 (38.86) and 360.84 (1557.19), respectively. On closer inspection, we also find that these errors vary little across counting stations (Figure \ref{figure:in_depth_SMAPE_b} and \ref{figure:in_depth_SMAPE_d}). This is a clear improvement over the multi-data-only model. Therefore, estimates predicted with sample counts and multi-source data are not only more accurate (around 2/3 lower) but also more reliable.

\section{Discussion}

Our research highlights the feasibility of estimating bicycle volumes for all streets across a city by leveraging open-source data together with long- and short-term counts and machine-learning models. Advances not only in data availability but also in analytical methods have made such purely data-driven approaches feasible.\\

We find that using already existing multi-source and long-term counting station data allows for predicting bicycle volume at unseen locations using XGBoost with a reasonable error for both daily values and annual averages. The most important of these data sources are crowdsourced Strava data, features indicating the time as well as infrastructure and socioeconomic indicators. Yet, the prediction error varies greatly between locations, which means that the model is able to predict certain streets very well and others much less so (with no apparent pattern). This is also anecdotally shown in the proof of concept, where the model performs well in capturing temporal trends and identifying high-volume areas, but shows shortcomings in reproducing intricate geographic nuances. We conclude that there may be unobserved and latent variables that remain unaccounted for despite our comprehensive inclusion of a wide range of features, which go beyond what has been done in the existing literature. Addtionally, collecting sample counts for unseen locations not only drastically reduces the error across all locations but also the variance across locations. The decrease is at around 2/3 on average. This experiment showed that spending resources to collect additional short-term counts may be worthwhile.\\ 

Municipal governments can replicate our model using data that are already owned by the city or can be obtained from third-party providers. Based on our findings, we advise that it is most relevant to obtain data from crowdsourced applications (Strava), infrastructure indicators from OpenStreetMaps, and, if available, socioeconomic data. We also recommend conducting multiple one-day sample counts at locations of interest to obtain more accurate results. Each day of sampling leads to a significant improvement in the estimate for that location. Using 10 days of sample data, our model provides policymakers with accurate and reliable estimates. Such estimates can allow them to make evidence-based decisions about infrastructure improvements or repairs. Busier roads can be prioritized, and financial expenditures can be justified by the number of cyclists they may benefit. Similarly, civil society can use such estimates to advocate for local infrastructure improvement needs.\\

In future research, more complex modeling approaches that take into account spatial and temporal dependencies can be another promising direction. Such approaches may also benefit from more ground truth data, particularly from more continuous counting stations to cover spatial variability. 
We hypothesize that more ground truth could further improve predictions for unseen locations and possibly reduce the need for sample data collection. Expanding the case study of Berlin to a comparative analysis with other cities could shed light on the generalizability of the approach.

\section{Methods}
\subsection{Data description \label{sec:methods_data_description}}
A table explaining each individual feature is included in the Appendix \ref{sec:appendix_detailed_features}, Table \ref{tabel:of_features}. In the following we elaborate on the data sources. All datasets are publicly available via the sources cited with one exception. Strava Metro provides their crowdsourced app data upon request \parencite{strava_metro_strava_2023}. 

\paragraph{Bicycle Counting Stations Data}
The Berlin city administration collects data on bicycle volume at various locations \parencite{senate_department_for_the_environment_mobility_consumer_and_climate_protection_berlin_zahlstellen_2023} (Appendix \ref{sec:appendix_overview_counting_stations} Figure \ref{table:overviewcounter}). The data come from long-term counting stations, which are permanent devices that identify passing bicycles through an electromagnetic field embedded in the ground. The city installed its first of 30 counting stations in 2012 and the most recent one in 2022. 10 of the stations are located on opposite sides of the street and also record the direction of flow, while in the other locations, there is only one counter for both directions. We sum counters on opposite sides of the same street into one count as we are interested in the number of bicycles passing by a certain location rather than their direction of flow. This reduces the number of counting locations to 20. Occasionally, counting stations are out of service (e.g., due to construction or malfunctioning), resulting in missing observations. We also exclude observations that are interpolated by the municipality, as the city does provide information on their interpolation method. 
Short-term bicycle counts have been conducted at 21 fixed locations repetitively on different dates by the city since 1983. We exclude short-term counters consisting of only one observation (one day) from the analysis. Consequently, the data set comprises information from a total of 12 short-term locations.
There is no publicly available information on the criteria used by the city to determine the placement of these counting stations, but upon closer examination, most are located closer to the city center and along high-traffic roads. A map of the counting stations, a table detailing the number of observations per station as well as some basic descriptions of the measurements are included in the Appendix \ref{sec:appendix_overview_counting_stations}.

\paragraph{Crowdsourced App Data}
We obtain crowdsourced app data from the Strava smartphone app, which allows users to track their speed, altitude gain, and exact route choice covered during physical activities such as cycling \parencite{strava_metro_strava_2023}.
Since Strava has made some of its data available for research and city administrations, other studies have used these data for a similar purpose (Table \ref{table:overview_factors_to_predict_cycling_volume}). 
Strava Metro has modified the data to protect individual privacy. The data includes only publicly available trip records (as opposed to trips taken in a private mode). They also do not provide individual trip information, but instead aggregate the trip counts into two formats. 
In the first format,  various features are available at the "street segment" level, which covers a street between two intersections. The data is available on an hourly basis. In the second format, features are available for regions in the form of hexagons, each spanning approximately 0.66km$^2$. In both formats bicycle counts are rounded to the nearest multiple of five, e.g., when 7 cyclists pass a street segment, Strava rounds it to 5. We use both the segment and the hexagon formats. For the segment data, we calculate the average of the available features for all segments within a certain radius of the counting station (500m, 1000m, 2000m, 5000m, whole city) at the daily level.  For the hexagon data, we include the features for both the hexagon in which the counter is located and the mean of the six surrounding hexagons. A graphical representation of the feature engineering process is included in the Appendix \ref{sec:appendix_feature_engineering_strava}. 
The Strava data is based only on the voluntary recording of Strava app users, and it has a sampling bias in its user base \parencite{lee_strava_2021}. Based on the few demographic indicators included in the data, a sampling bias towards young males with an ambitious riding style is apparent: 75.99\% of trips were recorded by male users, only 3.76\% of the trips were recorded by users aged 55 and over, e-bike trips contribute only 0.17\%, and the average speed is relatively high at 21.14km/h (see Appendix \ref{sec:comparison_descriptives_strava_bikesharing} for more details).
\paragraph{Bike-Sharing Data}
The bike-sharing data used in this study consists of individual trips from free-floating bike-share systems. In these systems, bicycles are available for pick up and return anywhere within the city, unlike systems dependent on designated stations for both pickup and drop-off. It comprises two distinct periods. The months from April to December 2019, covering the providers Call-a-bike and Nextbike, as provided to us by \cite{citylab_berlin_shared_2019}. And from June to December 2022, covering only Nextbike, which we collected ourselves via their application programmable interface (API) \parencite{nextbike_official_2020}. The data is made available for download \parencite{kaiser_bike-sharing_2023}. To the best of our knowledge, only \cite{miah_estimation_2022} use information on bike-sharing to predict cycling counting stations. These data provide details of individual trips, unlike crowdsourced data (Strava), which only offer aggregate counts.
The 2022 bike-sharing data was collected as follows. Nextbike's application programming interface furnishes real-time information on the location of all accessible bicycles, each identified by a unique bike ID, at a minute-by-minute granularity. When a bike is rented, it is temporarily removed from the available list and reappears when it is returned. By querying the list at one-minute intervals, we can accurately record trips to the minute, providing precise departure and arrival points and the respective times for every trip.  
For both bike-sharing datasets, 2019 and 2022, only the start and end points of each trip are available. We impute the route using OpenStreetMap as of July 2022 and the designated routing algorithm tailored for bicycles. OpenStreetMaps facilitates route planning for different modes of transportation, by adapting the suggested route according to the chosen mode \parencite{openstreetmap_contributors_planet_2017}. It is important to note that the resulting trajectories are approximations of the actual routes taken by bike-sharing users. 
Based on the routed trips, we perform data cleaning to account for possible incorrectly recorded journeys. We exclude trips shorter than 100m (0.64\% of total trips), which may be due to errors in  GPS measurements or aborted trips, e.g., due to a broken bike. Similarly, we exclude trips longer than the 45km diameter of Berlin (0.005\%), shorter than 120 seconds (1.63\%), and longer than 10 hours (6.05\%), assuming that incorrect use of the rental system is the cause. Finally, we exclude trips with an average speed slower than 2km/h (16.57\%) or faster than 40km/h (10.87\%). After cleaning, the data contain 1,333,737 bike-sharing trips.
We feature engineer the bike-sharing data based on the methodology proposed by \cite{miah_estimation_2022}. For each counter, we count the number of bikes passing, the number of bikes whose rental started, and the number of bikes whose rentals ended within a certain radius within a day. As radii, we consider 250m, 500m, 1000m, 2000m, 5000m, and the whole city. We provide an example of feature engineering for the bike-sharing data in the Appendix \ref{sec:feature_engineering_bike-sharing}.
The following limitations to the bike-sharing data remain. In Berlin, the bike-sharing market is diverse, with multiple providers. We were only able to acquire data from two providers. These two maintain a fleet of traditional bicycles, unlike other companies that primarily offer e-bikes. In addition, neither Nextbike nor Call-a-bike responded to requests for information or provided detailed information about their data. This lack of transparency raises concerns about the stability of bike IDs, potentially leading to the inclusion of fictitious rides in our dataset due to how we compute trips. 
Also, these data are potentially biased, as bike-sharing users differ from private bicycle users. We do not have demographic information about the users of Nextbike or Call-a-bike, but bike-sharing users tend to have higher incomes and education \parencite{fishman_bikeshare_2016}. 

\paragraph{Weather data}
Cyclists are more exposed to environmental conditions than motorized traffic users, which affects their comfort while cycling and, consequently, their decision to use a bicycle. Research shows that weather conditions can have both positive and negative effects on cycling, resulting from both immediate and delayed weather effects \parencite{miranda-moreno_weather_2011}. 
The data we use comes from the German Weather Service and is provided by Meteostat \parencite{meteostat_weathers_2022}. We include various features at daily granularity. 
The weather indicators are for all of Berlin, i.e., they are the same for all counting locations but vary over time.

\paragraph{Infrastructure data}
We include infrastructure data on road conditions, points of interest, and the land use around the counting stations.
Cycling and road infrastructure plays a critical role in increasing cyclists' perception of safety and, consequently, influence bicycle use \parencite{moller_cyclists_2008}. The infrastructure features also affect how many trips may be made to an area. Similarly, points of interest, such as schools and shops, can influence bicycle volumes at different times throughout the day and week \parencite{strauss_spatial_2013}.  
From \cite{openstreetmap_contributors_planet_2017}, we obtain information about the maximum speed allowed for motorized traffic, the type of bike lane at the exact location of the counting station, and the number of different points of interest within different radii (500m, 1000m, 2000m, and 5000m). We also compute the distance of the counting stations from the city center, following the definition of a city center used by OpenStreetMap. 
Data from the city of Berlin provides information on land use, such as for parks or industry, which can impact the timing and volume of human frequenting in various areas. This data is collected at the ``planning area'' level: For urban planning purposes, the city is divided into planning areas that represent neighborhoods; each planning area has an average size of about 2 km$^2$. The city collected the indicators in 2015 \parencite{berlin_open_data_nutzung_2022,  senate_department_for_urban_development_building_and_housing_lebensweltlich_2023}. We use data from the planning area surrounding each counting station. To standardize the measurements, we convert the features from square kilometers into percentages. For example, instead of stating that 0.5km$^2$ within the planning area surrounding the counting station is occupied by parks, we express it as 25\% occupied by parks. 

\paragraph{Socioeconomic data}
Bicycle use varies with regard to socioeconomic characteristics such as age and gender \parencite{goel_cycling_2022}.
We obtain socioeconomic data from the city of Berlin \parencite{berlin-brandenburg_office_of_statistics_kommunalatlas_2020}. The data is available at the level of ``planning areas'' (see the infrastructure data for details). For each counter, we use the indicators of the planning area in which it is located. Since the socioeconomic data are only available until 2020, we use the 2019 observations for the same year and the 2022 observations for 2020.  
Therefore, the data has spatial and temporal variation, but the data for 2022 remains an approximation. 

\paragraph{Motorized traffic data}
Similar to the bicycle counting stations, the city administration conducts counts of motorized traffic at 267 counting stations (for 2019 and 2022) \parencite{berlin_open_data_verkehrsdetektion_2022}.
To the best of our knowledge, we are unaware of any studies that have attempted to predict bicycle volume from motorized traffic counts. We hypothesized that this could be a valuable source of additional information as bicycle counts and motorized traffic counts may show similar patterns in terms of commuting peaks, weekday/weekend behavior, and locations of interest. 
The data includes the volume, type, and speed of motorized vehicles collected at various locations in Berlin. The detectors measure the features only on one side of the road (e.g., only eastbound traffic).
We compute the respective mean values of these motorized traffic features of all traffic counters within a 6-kilometer radius and for the city as a whole, all on a daily basis. The choice of a 6-kilometer radius as our spatial unit is intentional, as it is the smallest possible radius for the feature to be available for each counting station. 
The main drawback of the data is that the motorized traffic counting stations are unevenly distributed throughout the city (see Appendix \ref{sec:appendix_carcounter_location}). Therefore, not only do we have to employ a large radius, but we also have to compute the features for each counting station based on a different number of motorized traffic counting stations. 

\paragraph{Holiday and time data}
Traffic data can exhibit strong seasonality. We, therefore, include several time indicators: the day of the week, the day of the month, the month itself as features, the year, and whether it is a weekday. We also use features that indicate the presence of each public and school holiday \parencite{senate_department_for_education_youth_and_family_ferientermine_2022}.

\paragraph{Pre-processing of the combined data}
In the merged dataset, an observation represents a daily measurement from one counting station. We pre-process these data as follows: We drop a feature if it correlates more than 99\% with another feature, which is the case for the Strava non-e-bike trip count, which correlates with the Strava total trip count. We exclude infrastructure features that are constant across observations, as they do not provide additional predictive information. This is the case for the number of hospitals within 500m, the number of industries within 500m, 1000m, 2000m, and 5000m, and the percentage of land used for horticulture as they are all zero. The socioeconomic data is also missing for one counting station, which we replace with the mean of the respective features across all other counting stations.

\subsection{Data analysis \label{sec:data_analysis}}
\paragraph{Models and algorithms}
We implement all models with the Python library scikit-learn \parencite{pedregosa_scikit-learn_2011}. Based on the results, we select Extreme Gradient Boosting (XGBoost) as the best-performing model. It is an \acrshort{ml} algorithm that combines boosting and regularisation techniques. By iteratively adding decision trees to an ensemble model, it corrects the errors of the previous trees, resulting in a more robust and accurate model compared to standard decision trees or random forests. The trees are trained using a gradient descent optimization method, which updates the weights of the features to minimize a given loss function. In addition, the algorithm uses a technique called tree pruning to remove unnecessary leaves and nodes from the trees. 
For information on the other models, we would like to refer the reader to \textcite{geron_hands-machine_2022}.

\paragraph{Hyperparameter tuning}
For XGBoost, we tune the following hyperparameters with random search: the learning rate (controls the step size during the optimization process), the maximum depth of each tree (deeper trees can capture more complex relationships but can cause overfitting), the fraction of features used when constructing each tree (reducing overfitting also introducing randomness), the minimum sum of instance weight needed in a child (can prevent overfitting by controlling the minimum amount of instances required in each leaf) and a regularization parameter that encourages pruning of the tree \parencite{brownlee_xgboost_2016}. 
For the hyperparameters of the other models, we would like to refer the reader to the Appendix \ref{sec:appendix_models}.

\paragraph{Feature selection }
Given the richness of our features in the different data sources, we employ model-specific \acrlong{fs} (\acrshort{fs}), which allows for a reduction of computational needs and can improve the performance of the models. For each model, we test univariate \acrshort{fs} with SelectKBest, recursive feature elimination based on XGBoost, SelectFromModel with XGBoost, and \acrshort{fs} via Sequential Selection with Linear Regression \parencite{pedregosa_scikit-learn_2011}. The best feature selection for each model is assessed with \acrshort{logo} for both the 0h-24h and 7h-19h subsets and \acrshort{smape} and \acrshort{mae} separately.  Compared to dimensionality reduction, \acrshort{fs} has the additional advantage of allowing for feature importance analysis. Which \acrshort{fs} method is applied to which model is specified in the Appendix \ref{sec:appendix_feature_selection_methods}.

\paragraph{Error metrics}
For benchmarking, we choose two error metrics, \acrshort{mae} and \acrshort{smape}, which are defined as follows, with $n$ the number of observations, $y_i$ the true value, and $\hat{y}_i$ the prediction of the variable of interest:

\begin{equation}
    \mbox{MAE} = \frac{1}{n}\sum_{i=1}^{n}\lvert y_i-\hat{y}_i\rvert
\end{equation}

\begin{equation}
     \mbox{SMAPE} = \frac{1}{n} \sum_{i=1}^n \frac{|\hat{y}_i-y_i|}{(y_i+\hat{y}_i)/2}.
\end{equation}

We have chosen these metrics over the more commonly used error metrics: \acrlong{rmse} (\acrshort{rmse}), \acrlong{mape} (\acrshort{mape}), or standard errors based on the underlying distribution. The counting station measurements include several high outliers. Compared to the \acrshort{rmse}, the \acrshort{mae} places less emphasis on outliers, which is more suited to the right tail distribution. See Appendix \ref{sec:appendix_overview_counting_stations} for histograms and boxplots of the counts. We have chosen \acrshort{smape} over \acrshort{mape} to measure the relative error, as it yields small percentage errors when the true value is a high outlier. Additionally, \acrshort{smape} gives a symmetric measure that considers both overestimation and underestimation errors equally.

\paragraph{Grouped feature importance}
Computing feature importance for groups of features cannot be simply done by summing individual feature importance scores. Neither can one sum individual feature importance for tree-based methods. This method often overfits, boosting features that contribute little, and thus, they cannot be summed at the group level since this would overweight groups with many features \parencite{bramer_overfitting_2005, breiman_random_2001}. Nor can one sum up permutation-based features, as all features besides the one in question are known during the permutation, which does not sufficiently reveal the impact of a particular feature group when summed up \parencite{plagwitz_supporting_2022}. \\

Here, we use the grouped feature importance as introduced by \cite {plagwitz_supporting_2022}. The data is split in the sense of cross-validation into training and test data. On each fold, the following is computed: A model is trained on the training data. The test data is replicated a certain amount of times, and in each replication, the features belonging to a feature group in question are permuted. The model is then applied to the permuted test sets. The change in performance is estimated and averaged across all permuted test sets. This process is repeated for every feature group. The mean between the cross-validation folds returns the final grouped feature importance score, which provides the information gain per group.
These scores are not comparable across models but only within each model.

\paragraph{Sample weights in training}
We employ sample weights during model training with the sample data to enhance the model's emphasis on these particular observations. Sample weights assign different weights to individual observations in the training dataset. This is useful when dealing with imbalanced datasets or when certain samples are more critical than others. The latter is the case in our setting. In XGBoost, sample weights can be assigned to each instance, influencing the contribution of that instance's error to the overall loss function during training. This way, samples with higher weights contribute more to the model's updates, thus affecting the model's learning process \parencite{pedregosa_scikit-learn_2011}.\\


\newpage

\begin{appendix}
\section{Appendix}\label{appendixA}
\subsection{Overview counting stations} \label{sec:appendix_overview_counting_stations}
\begin{figure}[H]
    \centering
    \includegraphics[width=0.7\textwidth]{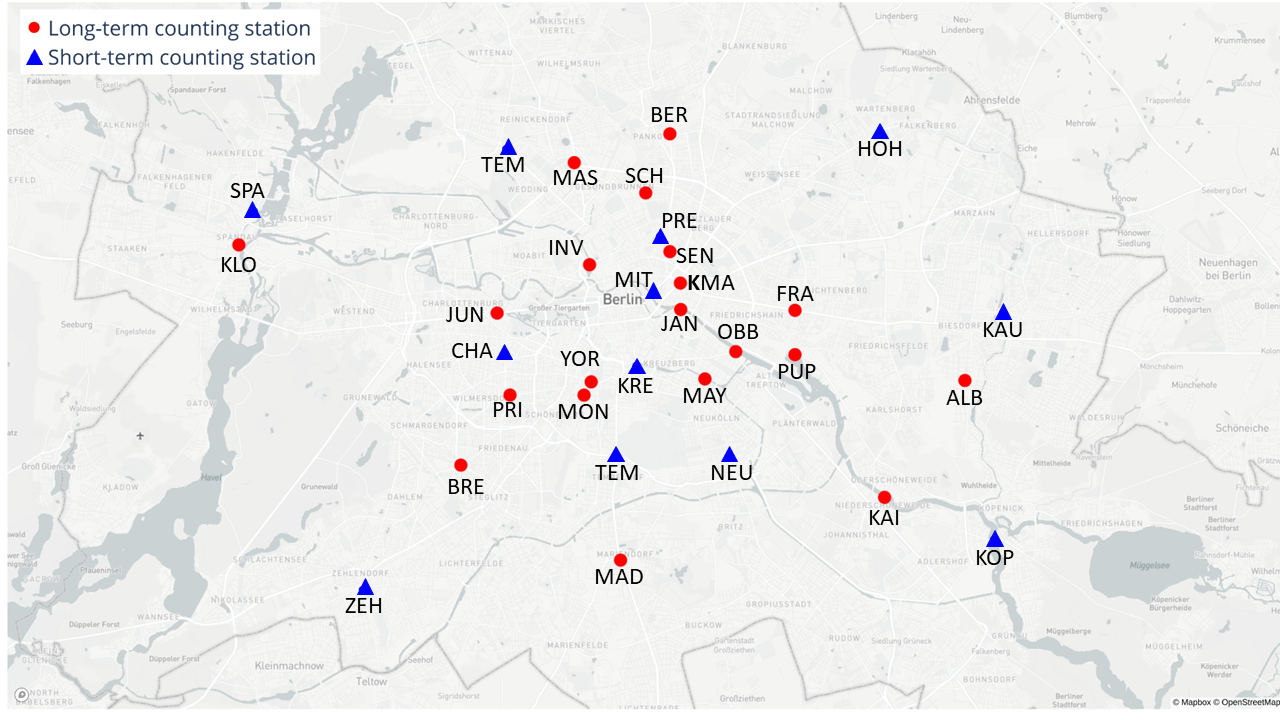}
    \caption{\textit{Location of the 12 short-term and 20 long-term counting stations within Berlin.}}
    \label{fig:CounterLocation}
\end{figure}

\begin{figure}[H]
    \begin{subfigure}{7.8cm}
      \centering\includegraphics[width=\textwidth]{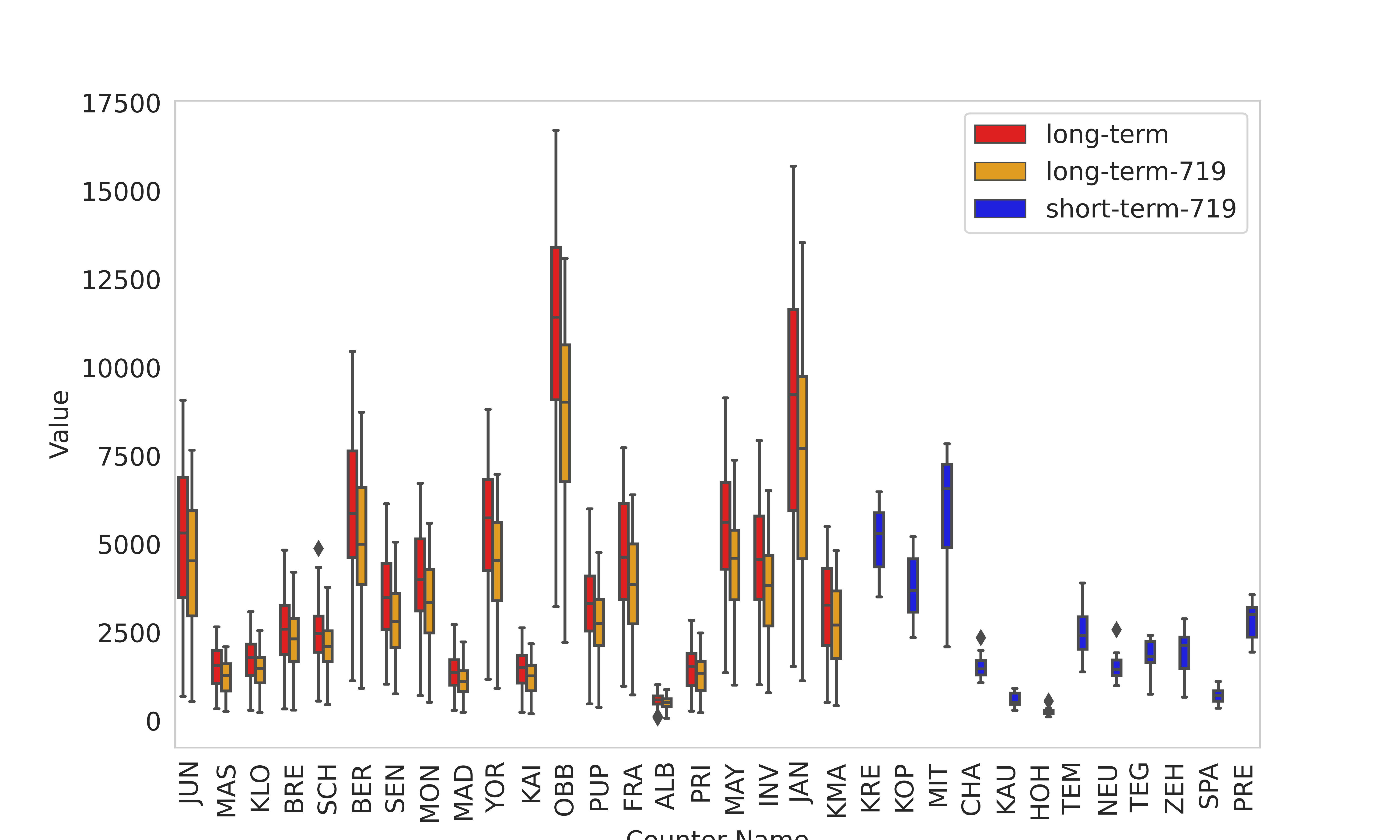}
    \caption{\textit{Bicycle volume per counting station}}
    \label{figure:boxplot_counters_daily}
  \end{subfigure}
    \begin{subfigure}{6.2cm}
    \centering\includegraphics[width=\textwidth]{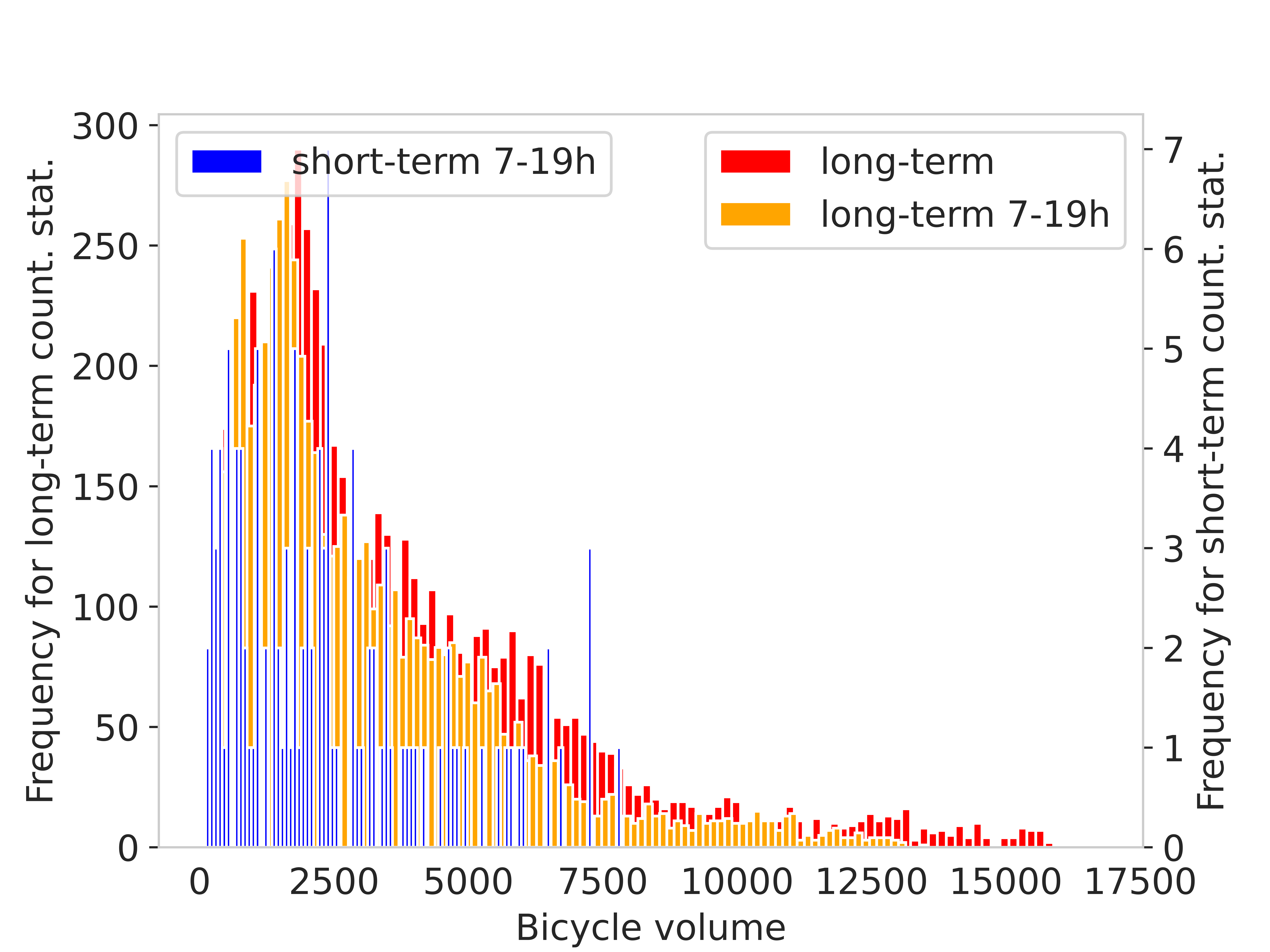}
    \caption{\textit{Bicycle volume across counting stations}}
    \label{figure:histogram_counters_daily}
  \end{subfigure}
\caption{\textit{Descriptive statistics of the counter stations' measurements,for the time period considered in this paper.}}
\label{fig:DescriptiveStatistics_Counters}
\end{figure}

This section provides further details on the counting stations, including an overview of the available counting stations (Table \ref{table:overviewcounter}) and a mapping of them (Figure \ref{fig:CounterLocation}), as well as some descriptions of the measurements (Figure \ref{fig:DescriptiveStatistics_Counters}). 

The boxplots and histograms of counting stations' measurements reveal distinct patterns. The boxplot (Figure \ref{figure:boxplot_counters_daily}) demonstrates that long-term stations have very infrequent outliers. Conversely, short-term stations show a lower mean count with few outliers. This disparity arises as they cover a shorter period, including fewer days with extreme events. Short-term stations consider only daytime measurements (7-19h), omitting the lower nighttime counts. This assumption is supported by Figure \ref{figure:histogram_counters_daily}, depicting permanently higher values for the long-term, in comparison to the long-term 7-19h. The distributions,  exhibit a right-skewed, long-tailed pattern, occasionally indicating notably high cycling volumes. However, the right-skewedness is less pronounced for short-term stations.\par

\begin{table}[H]
\tabcolsep=0pt%
\TBL{\caption{\textit{The long-term and short-term counting stations.}\label{table:overviewcounter}}}
{\begin{fntable}
\begin{tabular*}{\textwidth}{@{\extracolsep{\fill}}llcccc@{}}\toprule%
\TCH{Counter} & \TCH{Location} & \TCH{Two-way combined} & \TCH{ Installed in} & \TCH{No° 0h-24h} & \TCH{No° 7h-19h} \\
\TCH{name\smash{\footnotemark[1]} } & \TCH{} & \TCH{ to one-way\smash{\footnotemark[2]} } & \TCH{} & \TCH{ measurements\smash{\footnotemark[3]}} & \TCH{ measurements \smash{\footnotemark[3]}}
\\\midrule

\multicolumn{3}{l}{Long term counting stations (20 locations)}\\
 \midrule
JAN &                           Jannowitzbrücke &        True &            2015 &                       380 &                           381 \\
BRE &                          Breitenbachplatz &        True &            2016 &                       380 &                           381 \\
PRI &                       Prinzregentenstraße &       False &            2015 &                       380 &                           381 \\
FRA &                         Frankfurter Allee &        True &            2016 &                       381 &                           382 \\
  BER &                           Berliner Straße &        True &            2016 &                       379 &                           380 \\
SCH &                            Schwedter Steg &       False &            2012 &                       376 &                           378\\
MON &                          Monumentenstraße &       False &            2015 &                       373 &                           376\\
 MAY &                               Maybachufer &       False &            2016 &                       380 &                           381 \\
KAI &                                Kaisersteg &       False &            2016 &                       379 &                           380 \\
 MAS &                                Markstraße &       False &            2015 &                       376 &                           377 \\
MAD &                         Mariendorfer Damm &        True &            2016 &                       374 &                           376  \\
KLO &                             Klosterstraße &        True &            2016 &                       373 &                           374 \\
PUP &                    Paul-und-Paula-Uferweg &       False &            2015 &                       351 &                           352 \\
ALB &                            Alberichstraße &       False &            2015 &                       355 &                           356 \\
OBB &                            Oberbaumbrücke &        True &            2015 &                       218 &                           218 \\
INV &                           Invalidenstraße &        True &            2015 &                       204 &                           205 \\
YOR &                               Yorckstraße &        True &            2015 &                       179 &                           180 \\
KMA &                          Karl-Marx-Straße &       False &            2021 &                       164 &                           164 \\
JUN &                       Straße des 17. Juni &        True &            2021 &                       164 &                           164 \\
SEN &                           Senefelderplatz &       False &            2022 &                        53 &                            53 \\
        \midrule\multicolumn{3}{l}{Short term counting stations (11 locations)}\\
 \midrule
 CHA &    Joachimsthaler Str./ &        True &            2001 &                         0 &                            10 \\
    &    Lietzenburger Str. &        &            &                          &                             \\
TEM &                          Tempelhofer Damm &       False &            2011 &                         0 &                            10 \\
TEG &                         Scharnweberstraße &       False &            2011 &                         0 &                            10 \\
SPA &         Schönwalder Str. &        True &            2001 &                         0 &                            10 \\
 &         Neuendorfer Str. &         &            &                          &                             \\
KRE &                 Zossener Str./ &        True &            2001 &                         0 &                            10 \\
 &                 Blücherstr. &        &             &                          &                             \\
HOH &  Pablo-Picassso-Str./ &        True &            2001 &                         0 &                            10 \\
&   Falkenseer Chaussee &        &            &                          &                             \\
KOP &                              Lange Brücke &        True &            2001 &                         0 &                            10 \\
PRE &           Kastanienallee /  &        True &            2001 &                         0 &                            10 \\
 &           Schwedter Str. &         &             &                          &                             \\
NEU &                          Karl-Marx-Straße &       False &            2011 &                         0 &                            10 \\
MIT &       Karl-Liebknecht-Str./ &        True &            2001 &                         0 &                            10  \\
 &       Spandauer Str. &         &             &                          &                              \\
ZEH &          Teltower Damm /  &        True &            2001 &                         0 &                            10 \\
 &           Schönower Straße &         &            &                        &                            \\
KAU &                     Altentreptower Straße &       False &            2011 &                         0 &                             9 \\

\botrule
\end{tabular*}%
\footnotetext[1]{We use the abbreviation throughout the paper, to pinpoint the individual counters.}%
\footnotetext[2]{In some locations the counter stations count the passing bikes independently for the different sides of the street. In these cases, we combined them, ignoring the direction of the flow.}%
\footnotetext[3]{The number of observations refers to the time span in which all other necessary features were available.}%

\end{fntable}}
\end{table}

\subsection{Location of motorized traffic counting stations}\label{sec:appendix_carcounter_location}
\begin{figure}[!ht]
\begin{center}
\includegraphics[width=0.8\textwidth]{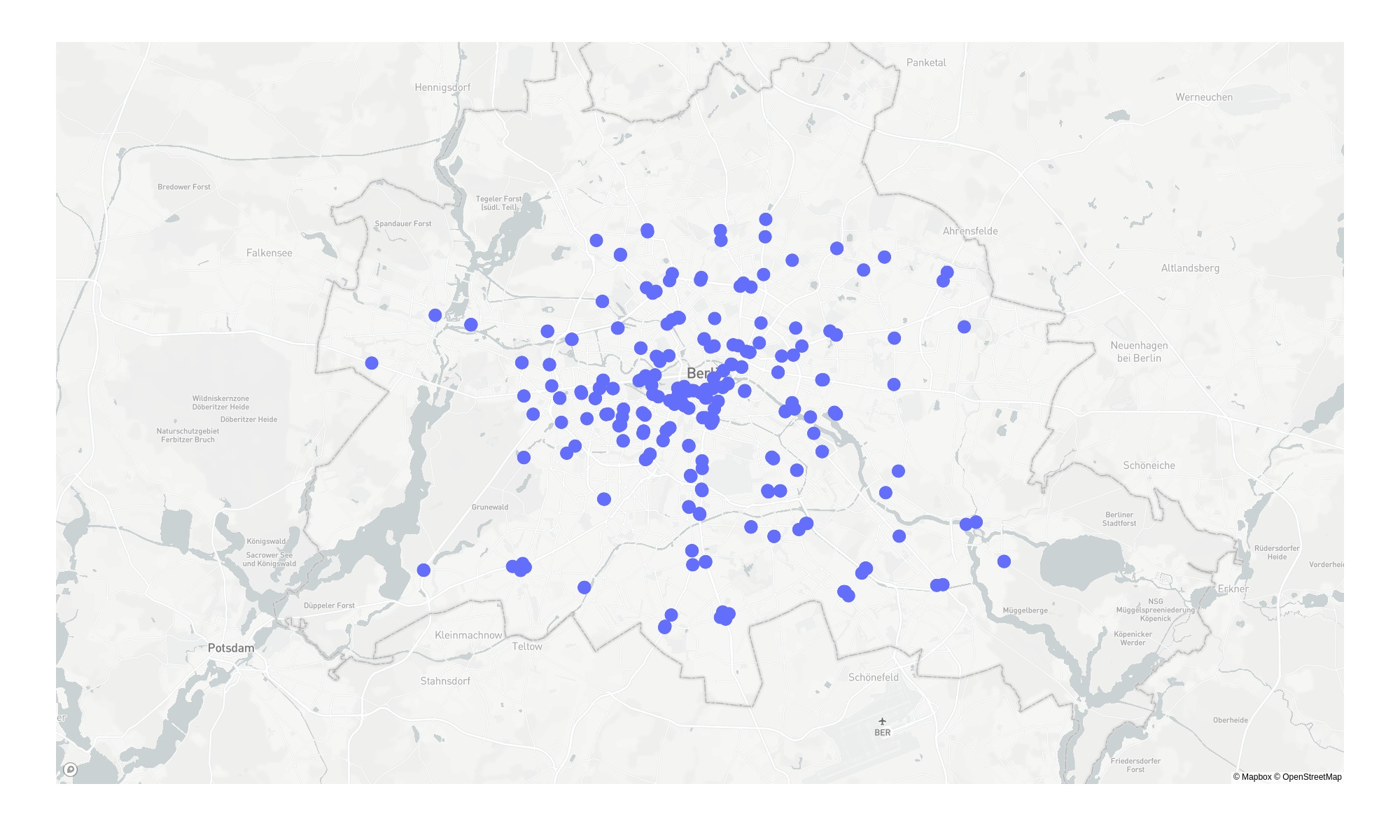}
\end{center}
\caption{\textit{Location of counting stations measuring motorized traffic.}} \label{fig:carcounter_location}
\end{figure} 

\subsection{Feature engineering crowdsourced data}
\label{sec:appendix_feature_engineering_strava}
\begin{figure}[H]
  \centering
    \subfloat[\textit{All segments lying, partially or fully, within a certain radius of the counting station (black) were considered for the feature engineering. Other segments were not considered (grey).}]{\includegraphics[width=0.49\textwidth]{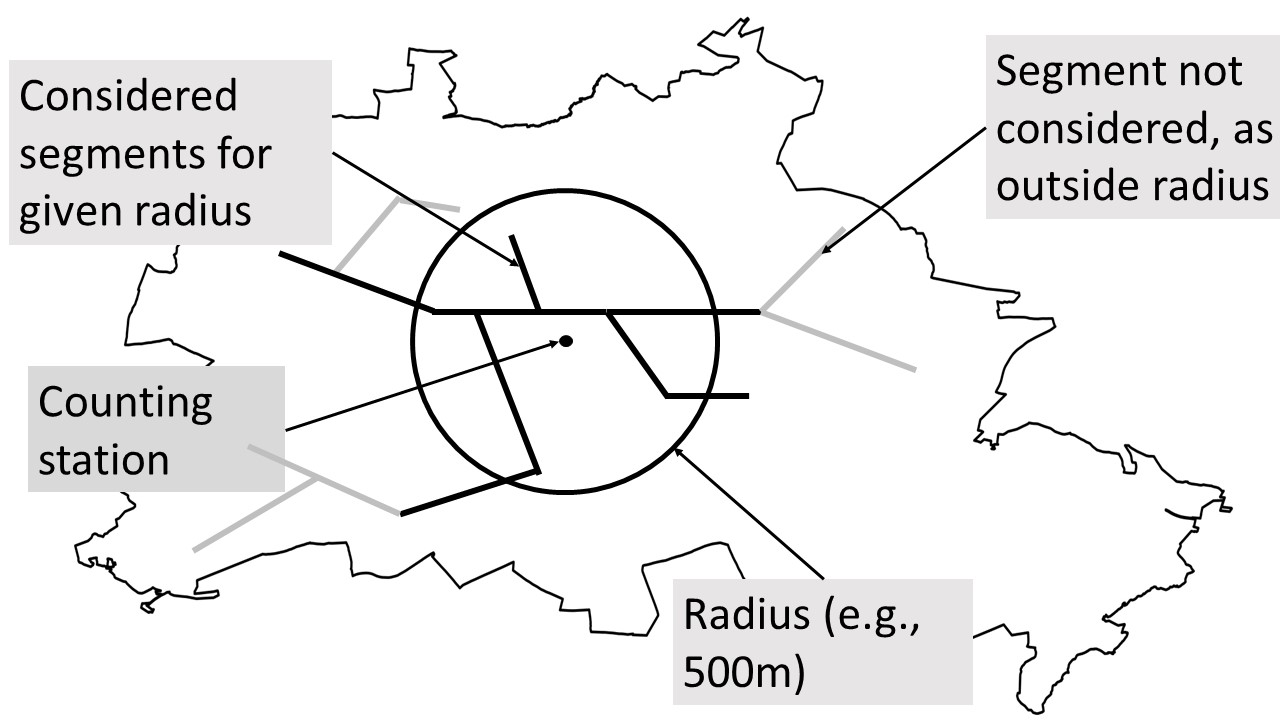}\label{fig:strava_featureengineering_segments}}
  \hfill
  \subfloat[\textit{The hexagon in which the counter is located (white, in the center) and its neighboring entities (grey).}]{\includegraphics[width=0.49\textwidth,  height=4cm]{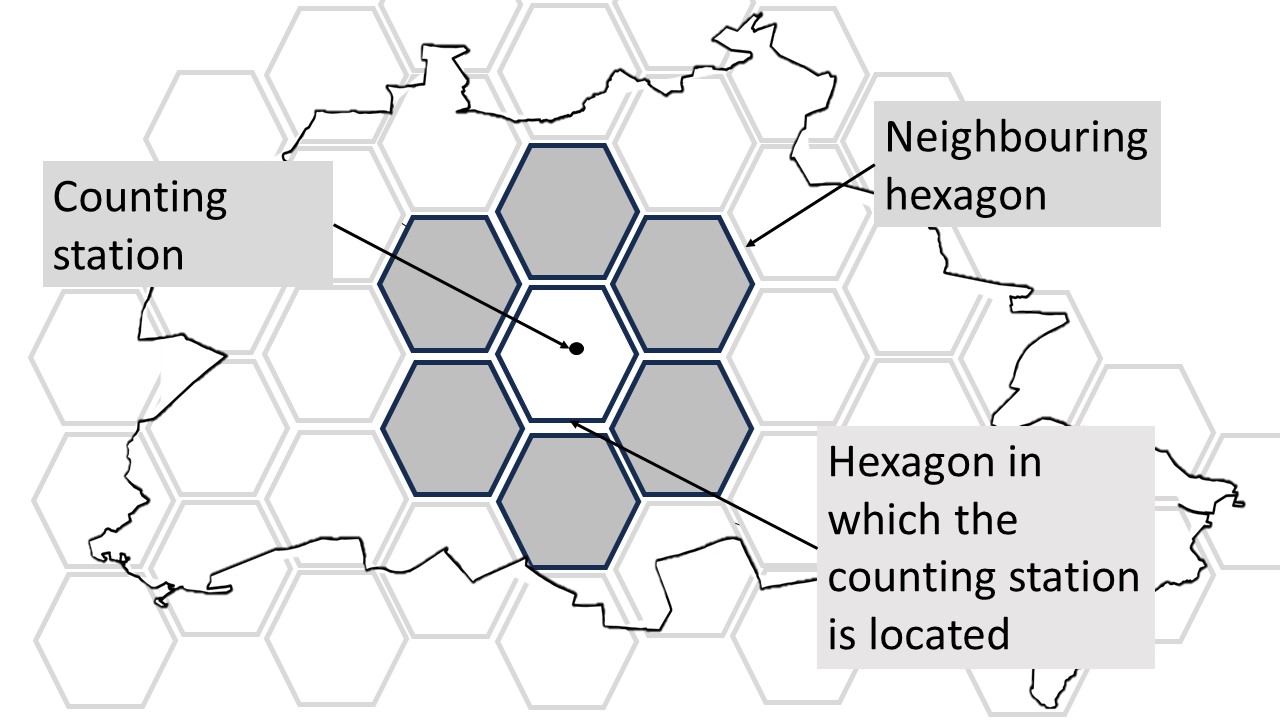}\label{fig:strava_featureengineering_hexagon}}
  \caption{\textit{The Strava data, both the hexagon and the street segment data, was feature-engineered. We computed the average across features for both data types, considering observations within a certain proximity. For the street segment data, we considered all segments within a certain radius. For the hexagon, we took the average of the features across the six neighboring hexagons. Additionally, we included the features for the hexagon, in which the relevant counting station is located.}}
\end{figure}

\newpage
\subsection{Feature engineering bike-sharing data}
\label{sec:feature_engineering_bike-sharing}

\begin{figure}[h]%
\FIG{\includegraphics[width=0.5\textwidth]{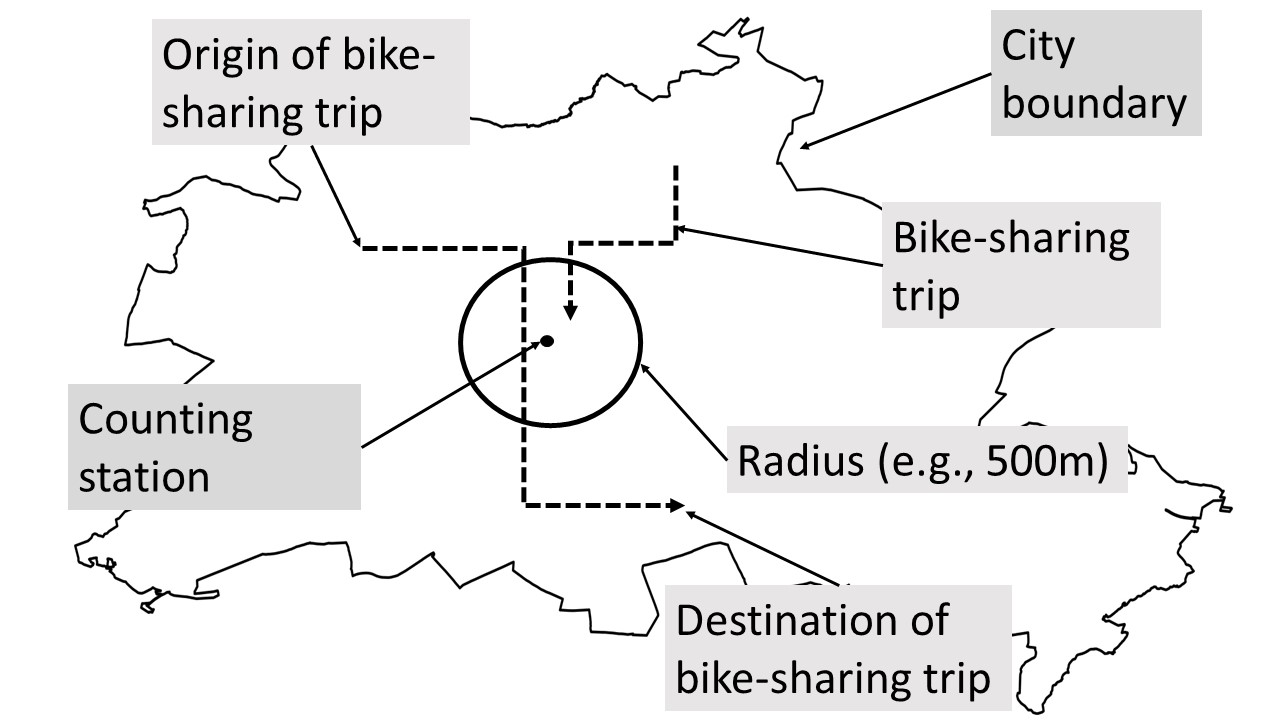}}
{\caption{\textit{The bike-sharing data was feature engineered based on a radius: For a given day all passing bike-sharing trips passing, starting or ending within a certain radius around the counting station in question were counted. This was also done for the entirety of the city. In this visualization, two bike-sharing trips are depicted. Given that both trips started and ended in a given day, the graph would produce a count of two passing, newly rented, and returned bike trips in the whole city, as well as two passing bike trips in the radius and one ending and zero originating. }}
\label{fig:featureengineering_bikesharing}}
\end{figure}
\newpage

\subsection{Comparison crowdsourced and bike-sharing data \label{sec:comparison_descriptives_strava_bikesharing}}

\begin{table}[H]
\tabcolsep=0pt%
\TBL{\caption{\label{table:descriptives_strava} \textit{Descriptives of the Strava and bike-sharing data. All specifications are in percent of the total trips recorded. The numbers indicate that bike-sharing trips are more evenly conducted throughout the whole day. Also, bike-sharing riders are much slower on average than Strava users, which seems reasonable, given the different quality of bike-sharing trips versus private bikes.}}}
{\begin{fntable}
\begin{tabular*}{\textwidth}{@{\extracolsep{\fill}}llll@{}}\toprule%

\TCH{ Group}           &    \TCH{Type}            &  \TCH{Strava}            & \TCH{Bike-sharing}  \\ \\\midrule
   Type of bike     & Non e-bike trips  &   97.49\%         & 100\%\\
                    & E-bike trips      &    0.17\%         & 0\%\\
   \midrule
   Purpose of trip\smash{\footnotemark[1]}  & Commute trips     &   39.95\%& NA\\
                    &Leisure trips      &   60.22\%         &NA\\
    \midrule
   Sex of user      & Male              &   75.99\%         &NA\\
                    &Female             &   10.69\%         &NA\\
                    &Gender unspecified &    0.66\%         &NA\\
    \midrule
    Age of user     &  18-34            &   25.99\%         &NA\\
                    & 35-54             &   47.97\%         &NA\\
                    & 55-65             &    3.70\%         &NA\\
                    &65+                &    0.06\%         &NA\\
    \midrule
    Time of trip\smash{\footnotemark[2]}    & Morning &   18.71\%         & 23.65\%\\
                    & Midday            &   20.94\%         &27.09\%\\
                    & Evening           &   34.31\%         &32.88\%\\
                    & Night             &    6.97\%         &16.38\%\\
    \midrule
    Basic parameters & Average distance & NA                & 2.90 km\\
                    & Average duration  &  NA               & 24.31 min\\
                    & Average speed     & 21.14km/h         & 11.05 km/h\\
\botrule
\end{tabular*}%
\footnotetext[1]{Is categorized by Strava. Strava identifies commutes through a model This model utilizes the "commute" tag provided by Strava members as ground truth. The term "commuting" encompasses all trips that are not related to leisure activities \parencite{strava_metro_strava_2023}}
\footnotetext[2]{ Morning: 5h - 10h, midday: 10h - 15h, evening: 15h - 20h, overnight: 20h - 5h. The Strava data was categorized by Strava. For the bike-sharing data, we considered the moment of departure.}%
\end{fntable}}
\end{table}

Usage patterns differ between Strava and bike-sharing. On average bike-sharing usage is more evenly distributed throughout the day, whereas Strava trips are more likely to be recorded during the midday and evening. Also, bike-sharers ride on average with a speed of 11.05 km/h whereas Strava are almost at double the speed with 21.14 km/h. This seems reasonable,  as Strava is used heavily to track sporting activities and bike-sharing bikes tend to be of lower quality.

\subsection{Detailed feature list \label{sec:appendix_detailed_features}}

\begin{landscape}
\footnotesize
\begin{longtable}{p{5cm}|p{2cm}|p{1.5cm}|p{1.5cm}|p{1cm}|p{1.5cm}|p{1.5cm}|p{3cm}}
 \caption{\label{tabel:of_features}\textit{Overview of all considered features}}
\\
\hline
Feature Name & Further explanation & Spatial scope & Timing used for daily model & No. of Features & Type & Scaling & Data Source \\ 
\hline
\multicolumn{5}{l}{Time Features}\\
\hline
Year & measurement from 2019 or 2022 & stationary & yearly& 1 & binary & & inherent \\
Month & indicating January through December & stationary & monthly & 1 & numerical & one-hot-encoded & inherent \\ 
Day of month & indicating numerical day of month& stationary & monthly & 1 & numerical & standardized & inherent \\ 
Weekday & indicating Monday through Sunday & stationary & daily & 1 & numerical & one-hot-encoded & inherent \\ 
Weekend & if Saturday or Sunday & stationary & daily  & 1 & binary & & inherent \\ 
 
\hline
\multicolumn{5}{l}{Vacation and holiday features}\\
\hline
School holiday & presence of school holiday & stationary & daily  & 1 & binary & & \parencite{senate_department_for_education_youth_and_family_ferientermine_2022} \\ 
Public holiday & presence of public holiday  & stationary & daily & 1 & binary & & \parencite{senate_department_for_education_youth_and_family_ferientermine_2022}\\
\hline
\multicolumn{5}{l}{Bike-sharing}\\
\hline
Originating/Returned/Rented & Number of & within a certain radius \footnotemark  to the counter & daily  & 15 & numerical & standardized & \parencite{nextbike_official_2020} and web scraped from Nextbike, as well as Call-a-bike \\
Originating/Returned/Rented & Number of & within the whole city & daily & 3 & numerical & standardized & \parencite{nextbike_official_2020} and web scraped from Nextbike, as well as Call-a-bike \\
\hline
\multicolumn{5}{l}{Strava}\\
\hline
No. of trips overall/originating/arriving/for leisure/ for commute/ morning/ midday/ evening/ overnight/ weekday/weekend & Number of & in the respective hexagon & daily & 18 & numerical & standardized & \parencite{strava_metro_strava_2023} \\
No. of trips overall/originating/arriving/for leisure/ for commute/ morning/ midday/ evening/ overnight/ weekday/weekend & Number of & in the six neighboring hexagons & daily & 18 & numerical & standardized &\parencite{strava_metro_strava_2023}   \\
No. of total trips/non-e-bikes/e-bikes/number of people/for commute/leisure/ sex (female, male and unspecified gender)/various age groups (18-34, 35-54, 55-64 and 65+), / morning/ midday/ evening/ overnight counted, average speed & Number of & in the segments within a certain radius \footnotemark[\value{footnote}] & daily & 44 & numerical & standardized & \parencite{strava_metro_strava_2023} \\
No. of total trips/non-e-bikes/e-bikes/number of people/for commute/leisure/ sex (female, male and unspecified gender)/various age groups (18-34, 35-54, 55-64 and 65+), / morning/ midday/ evening/ overnight counted in the segments, average speed  & Number of & in the whole city of Berlin & daily & 11 & numerical & standardized & \parencite{strava_metro_strava_2023} \\

\hline
\multicolumn{5}{l}{Infrastructure}\\
\hline
Counter within inner city limits & indicating whether the counter located within the "Ring" (inner city limits) & counting station location & stationary  & 1 & binary & & based on \parencite{openstreetmap_contributors_planet_2017} \\
Latitude \& Longitude &  & counting station location & stationary& 2 & numerical & standardized & \parencite{senate_department_for_the_environment_mobility_consumer_and_climate_protection_berlin_jahresdatei_2022} \\
Distance to city center & in km & counting station location & stationary & 1 & numerical & standardized & \parencite{openstreetmap_contributors_planet_2017} \\
Maximum speed & in km/h & counting station location & stationary & 1 & categorical & one-hot-encoded & \parencite{openstreetmap_contributors_planet_2017} \\
Bicycle lane type &  & counting station location & stationary & 1 & categorical & one-hot-encoded & \parencite{openstreetmap_contributors_planet_2017} \\
No. of shops &  & within a certain radius \footnotemark[\value{footnote}] & stationary  & 4 & numerical & standardized & \parencite{openstreetmap_contributors_planet_2017} \\
No. of education &  & within a certain radius \footnotemark[\value{footnote}] & stationary  & 4 & numerical & standardized & \parencite{openstreetmap_contributors_planet_2017} \\
No. of hotels &  & within a certain radius \footnotemark[\value{footnote}] & stationary  & 4 & numerical & standardized & \parencite{openstreetmap_contributors_planet_2017} \\
No. of hospitals &  & within a certain radius \footnotemark[\value{footnote}] & stationary  & 4 & numerical & standardized & \parencite{openstreetmap_contributors_planet_2017} \\
Percent of area used for farming &  & in the planning area & stationary & 1 & numerical & standardized & \parencite{senate_department_for_urban_development_building_and_housing_lebensweltlich_2023} \\
Percent of area used for horticulture &  & in the planning area & stationary  & 1 & numerical & standardized & \parencite{senate_department_for_urban_development_building_and_housing_lebensweltlich_2023} \\
Percent of area used for cemeteries &  & in the planning area & stationary & 1 & numerical & standardized & \parencite{senate_department_for_urban_development_building_and_housing_lebensweltlich_2023} \\
Percent of area used for waterways &  & in the planning area & stationary & 1 & numerical & standardized & \parencite{senate_department_for_urban_development_building_and_housing_lebensweltlich_2023} \\
Percent of area used for industry &  & in the planning area & stationary  & 1 & numerical & standardized & \parencite{senate_department_for_urban_development_building_and_housing_lebensweltlich_2023} \\
Percent of area used for private gardening &  & in the planning area & stationary  & 1 & numerical & standardized & \parencite{senate_department_for_urban_development_building_and_housing_lebensweltlich_2023} \\
Percent of area used for parks &  & in the planning area & stationary & 1 & numerical & standardized & \parencite{senate_department_for_urban_development_building_and_housing_lebensweltlich_2023} \\
Percent of area used for traffic areas &  & in the planning area & stationary& 1 & numerical & standardized & \parencite{senate_department_for_urban_development_building_and_housing_lebensweltlich_2023} \\
Percent of area used for forests &  & in the planning area & stationary & 1 & numerical & standardized & \parencite{senate_department_for_urban_development_building_and_housing_lebensweltlich_2023} \\
Percent of area used for residential housing &  & in the planning area  & stationary & 1 & numerical & standardized & \parencite{senate_department_for_urban_development_building_and_housing_lebensweltlich_2023} \\
\hline
\multicolumn{5}{l}{Socioeconomic indicators}\\
\hline
Population density & Inhabitants/km$^2$ & in the planning area & yearly & 1 & numerical & standardized & \parencite{senate_department_for_urban_development_building_and_housing_lebensweltlich_2023, berlin-brandenburg_office_of_statistics_kommunalatlas_2020} \& \parencite{senate_department_for_urban_development_building_and_housing_lebensweltlich_2023, berlin-brandenburg_office_of_statistics_kommunalatlas_2020} \\
Total number of inhabitants &  & in the planning area & yearly& 1 & numerical & standardized & \parencite{senate_department_for_urban_development_building_and_housing_lebensweltlich_2023, berlin-brandenburg_office_of_statistics_kommunalatlas_2020}\\
Average age & & in the planning area & yearly & 1 & numerical & standardized & \parencite{senate_department_for_urban_development_building_and_housing_lebensweltlich_2023, berlin-brandenburg_office_of_statistics_kommunalatlas_2020}\\
Gender distribution &  & in the planning area & yearly & 1 & numerical & standardized & \parencite{senate_department_for_urban_development_building_and_housing_lebensweltlich_2023, berlin-brandenburg_office_of_statistics_kommunalatlas_2020} \\
Share of population with migration background &  & in the planning area & yearly & 1 & numerical & standardized & \parencite{senate_department_for_urban_development_building_and_housing_lebensweltlich_2023, berlin-brandenburg_office_of_statistics_kommunalatlas_2020} \\
Share of foreigners (total, EU-foreigners, non-EU-foreigners) &  & in the planning area & yearly  & 3 & numerical & standardized & \parencite{senate_department_for_urban_development_building_and_housing_lebensweltlich_2023, berlin-brandenburg_office_of_statistics_kommunalatlas_2020}\\
Share of population unemployed &  & in the planning area & yearly & 1 & numerical & standardized & \parencite{senate_department_for_urban_development_building_and_housing_lebensweltlich_2023, berlin-brandenburg_office_of_statistics_kommunalatlas_2020} \\
Share of population with tenure exceeding 5 years &  & in the planning area & yearly  & 1 & numerical & standardized & \parencite{senate_department_for_urban_development_building_and_housing_lebensweltlich_2023, berlin-brandenburg_office_of_statistics_kommunalatlas_2020}\\
Net migration rate (moving to/away from the area) &  & in the planning area & yearly  & 1 & numerical & standardized & \parencite{senate_department_for_urban_development_building_and_housing_lebensweltlich_2023, berlin-brandenburg_office_of_statistics_kommunalatlas_2020}\\
Age-specific demographic proportions (individuals aged < 18 \& > 65) & & in the planning area & yearly & 2 & numerical & standardized & \parencite{senate_department_for_urban_development_building_and_housing_lebensweltlich_2023, berlin-brandenburg_office_of_statistics_kommunalatlas_2020} \\
Greying index &  & in the planning area & yearly & 1 & numerical & standardized & \parencite{senate_department_for_urban_development_building_and_housing_lebensweltlich_2023, berlin-brandenburg_office_of_statistics_kommunalatlas_2020} \\
Birth rate &  & in the planning area & yearly  & 1 & numerical & standardized & \parencite{senate_department_for_urban_development_building_and_housing_lebensweltlich_2023, berlin-brandenburg_office_of_statistics_kommunalatlas_2020} \\
\hline
\multicolumn{5}{l}{Weather}\\
\hline
Average temperature & in °C & stationary & daily & 1 & numerical & standardized & \parencite{meteostat_weathers_2022} \\
Daily maximum temperature & in °C & stationary & daily  & 1 & numerical & standardized & \parencite{meteostat_weathers_2022} \\
Daily minimum temperature & in °C & stationary & daily  & 1 & numerical & standardized & \parencite{meteostat_weathers_2022} \\
Precipitation & in mm & stationary & daily & 1 & numerical & standardized & \parencite{meteostat_weathers_2022} \\
Maximum snow depth & in mm & stationary & daily & 1 & numerical & standardized & \parencite{meteostat_weathers_2022} \\
Sunshine duration & in minutes & stationary & daily & 1 & numerical & standardized & \parencite{meteostat_weathers_2022} \\
Average wind speed & in km/h & stationary & daily & 1 & numerical & standardized & \parencite{meteostat_weathers_2022} \\
Wind direction & in degrees & stationary & daily  & 1 & numerical & standardized & \parencite{meteostat_weathers_2022} \\
Peak wind gust & in km/h & stationary & daily & 1 & numerical & standardized & \parencite{meteostat_weathers_2022} \\
Average sea-level air pressure & in hPa & stationary & daily  & 1 & numerical & standardized & \parencite{meteostat_weathers_2022} \\
\hline
\multicolumn{5}{l}{Motorized Traffic}\\
\hline
Total no° of vehicles / cars / lorries  &  & within a 6km radius to the counter & daily & 3 &numerical&   standardized & \parencite{berlin_open_data_verkehrsdetektion_2022}\\
Total no° of vehicles / cars / lorries  &  & within the whole city & daily & 3 & numerical & standardized & \parencite{berlin_open_data_verkehrsdetektion_2022}\\
Speed of vehicles / cars / lorries  &  & within a 6km radius to the counter  & daily & 3& numerical  & standardized & \parencite{berlin_open_data_verkehrsdetektion_2022} \\
Speed of vehicles / cars / lorries  &   & within the whole city & daily  & 3& numerical & standardized & \parencite{berlin_open_data_verkehrsdetektion_2022} \\
 \bottomrule
\end{longtable}
\footnotetext{The features are each computed for a radius of 0.5, 1, 2, and 5km.}   \enspace \enspace    

\end{landscape}

\subsection{ML models' hyperparameters\label{sec:appendix_models}}

\begin{table}[H]
\tabcolsep=0pt%
\TBL{\caption{\textit{Models and their tuned hyperparameters.\label{table:hyperparameters}}}}
{\begin{fntable}
\begin{tabular*}{\textwidth}{@{\extracolsep{\fill}}ll@{}}\toprule%
\TCH{Models} & \TCH{Tuned hyperparameters}\\\midrule
 Linear Regression & -\\  
 Decision Tree      & maximum depth, minimum samples for splitting, min samples in leaf \\ 
                    &  node, splitting criterion \\  
 Gradient Boosting  & number of estimators, learning rate, maximum depth, minimum samples \\  
                    &  for splitting, minimum samples in leaf node\\  
 XGBoost            & learning rate, maximum depth, subsample size, column subsampling  \\  
                    &rate, minimum child weight, gamma \\ 
 Random Forest      & number of estimators, maximum depth, minimum samples for splitting, \\
                    &   bootstrap sampling\\
  Support Vector Regression & C (regularization parameter), kernel choice, degree (for polynomial  \\  
         & kernel), gamma, epsilon \\  
 Shallow Neural Network & hidden layer sizes, activation function, learning rate\\
\botrule
\end{tabular*}%
\end{fntable}}
\end{table}

\subsection{Feature selection methods \label{sec:appendix_feature_selection_methods}}

\begin{table}[H]
\tabcolsep=0pt%
\TBL{\caption{\textit{Models and the applied feature selection method.\label{tale:feature_selection}}}}
{\begin{fntable}

\begin{tabular*}{\textwidth}{@{\extracolsep{\fill}}ll@{}}\toprule%
\TCH{Models} & \TCH{Feature Selection }\\\midrule
 Linear Regression & Recursive Feature Elimination with Linear Regression\\  
 Decision Tree      & Recursive Feature Elimination with Linear Regression \\  
 Gradient Boosting  & Feature Selection via Sequential Feature Selection XGBoost\\  
 XGBoost            & Recursive Feature Elimination with Linear Regression (MAE),\\
                    & Feature Selection via Sequential Feature Select. with XGBoost (SMAPE) \\ 
 Random Forest      & Recursive Feature Elimination\\
  Support Vector Regression & Regression Recursive Feature Elimination with Linear Regression \\  
 Shallow Neural Network & Network Recursive Feature Elimination with Linear Regression\\
\botrule
\end{tabular*}%
\end{fntable}}
\end{table}

\end{appendix}

\newpage

\section*{Abbreviations}

\noindent\textbf{AADB} \hspace{0.3cm}average annual daily bicycle volume.\\

\noindent\textbf{CV}\hspace{0.3cm} cross validation.\\

\noindent\textbf{FS}\hspace{0.3cm} feature selection.\\

\noindent\textbf{GPI}\hspace{0.3cm} Grouped Permutation Importance.\\

\noindent\textbf{LOGO}\hspace{0.3cm} leave-one-group-out.\\

\noindent\textbf{MAE} \hspace{0.3cm}mean absolute error.\\

\noindent\textbf{MAPE} \hspace{0.3cm}mean absolute percentage error.\\

\noindent\textbf{ML} \hspace{0.3cm}machine learning.\\

\noindent\textbf{RMSE}\hspace{0.3cm} root mean squared error.\\

\noindent\textbf{SMAPE} \hspace{0.3cm}symmetric mean absolute percentage error.\\

\begin{Backmatter}
\vspace{1cm}
\paragraph{Acknowledgments}
We are grateful to CityLab Berlin for providing their bike-sharing data. 

\paragraph{Funding Statement}
This research was supported by grants from the European Union’s Horizon Europe research and innovation program under Grant Agreement No 101057131, Climate Action To Advance HeaLthY Societies in Europe (CATALYSE). Furthermore, the authors acknowledge support through the Emmy Noether grant KL 3037/1-1 of the German Research Foundation (DFG). The funders had no role in study design, data collection and analysis, decision to publish, or preparation of the manuscript.

\paragraph{Competing Interests}
The authors declare no competing interests exist.


\paragraph{Ethical Standards}
The research meets all ethical guidelines, including adherence to the legal requirements of the study country (Germany).

\paragraph{Author Contributions}
Conceptualization \& Methodology: S.K., L.K.; Formal analysis, Investigation: S.K.; Resources: S.K., L.K., N.K.; Writing - Original Draft: S.K., L.K.; Writing - Review \& Editing: S.K., L.K., N.K.
All authors approved the final submitted draft.

\newpage
\printbibliography[heading=bibintoc,title={References}]
\end{Backmatter}

\end{document}